\documentclass[reprint,prb,amsmath,amssymb,amsfonts,superscriptaddress,floatfix,showpacs]{revtex4-1}
\usepackage{hyperref}
\usepackage[english]{babel}
\usepackage{braket}
\usepackage{bbm}
\usepackage{graphicx}
\usepackage{color}

\newcommand{\tr}{\operatorname{Tr}}
\newcommand{\ec}{\ensuremath{\mathrm{e}}}
\newcommand{\ic}{\ensuremath{\mathrm{i}}}

\newcommand{\Fig}[1]{Fig.~\ref{#1}}
\newcommand{\Sec}[1]{Sec.~\ref{#1}}
\newcommand{\Eq}[1]{Eq.~\eqref{#1}}

\begin{document}

\title{Matrix product state renormalization}

\author{M.~\surname{Bal}}
\affiliation{Department of Physics and Astronomy, Ghent University, Krijgslaan 281-S9, B-9000 Gent, Belgium}
\author{M.~M.~\surname{Rams}}
\affiliation{Institute of Physics, Jagiellonian University, \L{}ojasiewicza 11, 30-348 Krak\'ow, Poland}
\author{V.~\surname{Zauner}}
\affiliation{Vienna Center for Quantum Technology, University of Vienna, Boltzmanngasse 5, 1090 Vienna, Austria}
\author{J.~\surname{Haegeman}}
\affiliation{Department of Physics and Astronomy, Ghent University, Krijgslaan 281-S9, B-9000 Gent, Belgium}
\author{F.~\surname{Verstraete}}
\affiliation{Department of Physics and Astronomy, Ghent University, Krijgslaan 281-S9, B-9000 Gent, Belgium}
\affiliation{Vienna Center for Quantum Technology, University of Vienna, Boltzmanngasse 5, 1090 Vienna, Austria}

\begin{abstract}
The truncation or compression of the spectrum of Schmidt values is inherent to the matrix product state (MPS) approximation of one-dimensional quantum ground states. We provide a renormalization group picture by interpreting this compression as an application of Wilson's numerical renormalization group along the imaginary time direction appearing in the path integral representation of the state. The location of the physical index is considered as an impurity in the transfer matrix and static MPS correlation functions are reinterpreted as dynamical impurity correlations. Coarse-graining the transfer matrix is performed using a hybrid variational ansatz based on matrix product operators, combining ideas of MPS and the multi-scale entanglement renormalization ansatz. Through numerical comparison with conventional MPS algorithms, we explicitly verify the impurity interpretation of MPS compression, as put forward by V. Zauner {\it et~al.} [New J. Phys. {\bf 17}, 053002 (2015)] for the transverse-field Ising model. Additionally, we motivate the conceptual usefulness of endowing MPS with an internal layered structure by studying restricted variational subspaces to describe elementary excitations on top of the ground state, which serves to elucidate a transparent renormalization group structure ingrained in MPS descriptions of ground states.
\end{abstract}


\maketitle

\section{Introduction} In recent years, tensor network states have emerged as powerful theoretical and computational tools to investigate strongly correlated quantum many-body systems. By focusing on states rather than Hamiltonians, tensor networks are purposely designed in order to capture the entanglement structure inherent in physically relevant quantum states. For gapped quantum systems, matrix product states (MPS) \cite{fannes1992fcs,ostlund1995tldmr} are known to faithfully represent ground states in one spatial dimension \cite{verstraete2006faithfully}, and a plethora of numerical algorithms exist to variationally optimize MPS over the manifold of low-energy states of local Hamiltonians \cite{white1992density,vidal2004efficient,schollwock2005density,verstraete2008matrix,haegeman2011tdvp,haegeman2014unifying}. While local quantities can be approximated to very high precision, long-range behavior is not necessarily captured as accurately due to the exponential decay of correlations in MPS with finite bond dimension. This property is of particular importance in the context of quantum phase transitions, where it has fostered studies of finite entanglement scaling at criticality \cite{tagliacozzo2008scaling,pollmann2009theory,lauchli2013operator,stojevic2015conformal}.

The multi-scale entanglement renormalization ansatz (MERA) \cite{vidal2007entanglement} is an altogether different kind of tensor network tailored to the description of scale invariant systems. By introducing unitary disentangling operators, its layered structure is able to accommodate a proper and sustainable renormalization group (RG) flow along the intrinsic scale dimension of the network, even at criticality. Recently, it has been shown that MERA can be reinterpreted as stemming from a novel tensor network renormalization (TNR) scheme for coarse-graining two-dimensional tensor networks \cite{evenbly2014tnr,evenbly2015mera,evenbly2015local}. Unlike MPS, the MERA incorporates an explicit scale dimension into its network structure, which renders its real-space coarse-graining properties particularly explicit. Motivated by the initial discovery of remarkable similarities between holography and MERA \cite{swingle2012entanglement}, a lot of ongoing work \cite{hartman2013time,qi2013ehm,nozaki:2012holo,mollabashi:2013hologeom,miyaji:2015sscorr,bao:2015consistent,pastawski:2015qec,czech:2015igholo} has been concerned with clarifying the extent to which it is able to realize a true lattice version of the holographic principle.

It is well known that the entanglement content of a translation invariant MPS is entirely contained in the dominant eigenvector of its transfer matrix. In a recent publication \cite{zauner2015transfer}, it was observed that the other eigenvalues of the MPS transfer matrix contain additional useful information on the elementary excitations and corresponding dispersion relation of the system, providing an intriguing connection between the MPS transfer matrix and the spectral properties of the Hamiltonian. Put differently, there appears to be a highly non-trivial relationship between the excitation spectrum of a local translation invariant Hamiltonian and the local information and static correlations present in its ground state.

\begin{figure}[t]
 \centering
 \includegraphics[width=0.9\linewidth,keepaspectratio=true]{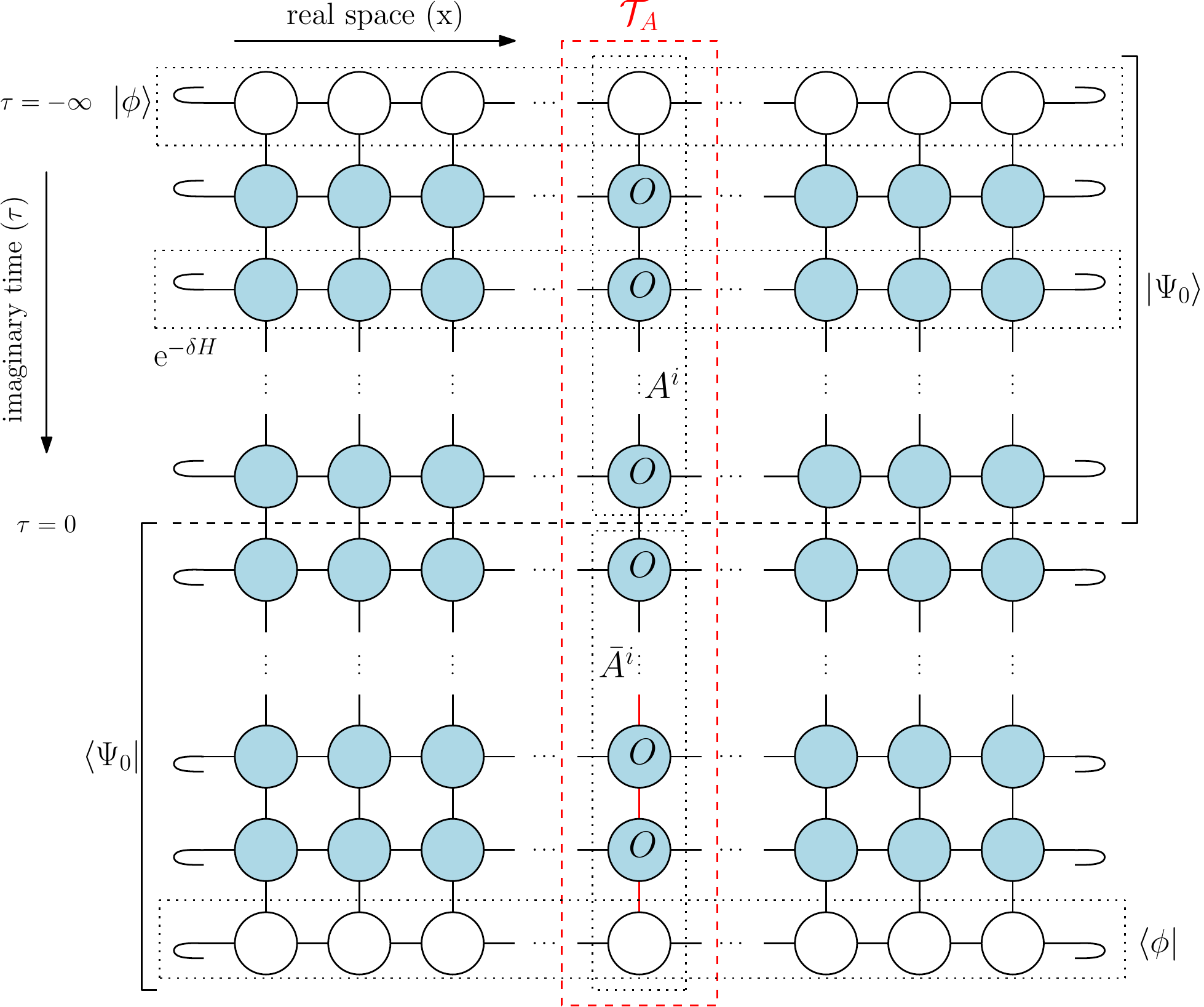} 
 \caption{(Color online) Two-dimensional tensor network representation of the Euclidean path integral corresponding to the ground state of a local one-dimensional translation invariant lattice Hamiltonian $H$. Horizontal slices are translation invariant matrix product operators built from a local tensor $O$ (blue circles), and correspond to imaginary time evolution with $\ec^{-\delta H}$. The MPS ground state $\ket{\Psi_0}$ is obtained by successively applying $\ec^{-\delta H}$ onto an initial MPS  $\ket{\phi}$. Vertical columns can then be interpreted as the MPS transfer matrix $\mathcal{T}_{A} = \sum_{i} A^{i} \otimes \bar{A}^{i}$, where $A^{i}$ and $\bar{A}^{i}$ respectively correspond to the translation invariant ket and bra MPS matrices.}
 \label{fig:pathintegral}
\end{figure}

In order to gain a better understanding of the origins of the MPS transfer matrix, it was proposed in Ref.~\onlinecite{zauner2015transfer} to treat the physical spin of a MPS as an impurity in the two-dimensional tensor network arising from the Euclidean path integral representation of the ground state. As depicted in \Fig{fig:pathintegral}, the transfer matrix of an infinite bond dimension MPS can then be identified with the exact quantum transfer matrix at zero temperature. By swapping the interpretation of (Euclidean) time and space, static correlation functions in the MPS are seen to correspond to temporal impurity correlations. The truncated finite-dimensional MPS transfer matrix obtained from numerical simulations will therefore contain only the degrees of freedom that are relevant to the impurity dynamics up to some infrared cutoff. This perspective suggests that the state compression involved in variational MPS techniques can be interpreted as an application of Wilson’s numerical renormalization group (NRG) \cite{wilson1975nrg} along the virtual (imaginary time) dimension of the system.

In this paper, we propose a variational ansatz based on matrix product operators (MPOs) to explicitly coarse-grain transfer matrices and set out to numerically verify the impurity picture proposed in Ref.~\onlinecite{zauner2015transfer}, beyond the analytical results obtained for the \textit{XY} model in Ref.~\onlinecite{rams2014truncating}. Having a layered decomposition of a MPS ground state at our disposal, we benchmark our method on the Ising model, propose an ansatz for the structure of MPS fixed point reduced density matrices, and study the effect of restricting a variational MPS ansatz for elementary excitations to a subspace of variational parameters. We furthermore translate our ansatz to the setting of free fermions, which allows us to further corroborate our findings by exploiting the additional free fermionic structure.

\section{MPS transfer matrices} Consider a one-dimensional lattice model and a local translation invariant Hamiltonian $H=\sum_{n \in \mathbb{Z}} h_{n,n+1}$ restricted to nearest-neighbour interactions. The translation invariant ground state of this system can be described by a uniform MPS 
\begin{align}
	\ket{\Psi[A]} = \sum_{{i}=1}^{d} \mathbf{v}_{L}^{\dagger} \left( \prod_{n \in \mathbb{Z}} A^{i_{n}} \right) \mathbf{v}_{R} \ket{\mathbf{i}}, \label{eq:mps}
\end{align}
where $\mathbf{v}^{\dagger}_{L}$ and $\mathbf{v}_{R}$ denote boundary vectors irrelevant for bulk properties. The state is hence completely determined by specifying a single tensor $A^{i} \in \mathbb{C}^{D \times d \times D}$, where $D$ and $d$, respectively, denote the bond dimension of the virtual level and the physical dimension of the local Hilbert space associated to each lattice site. To each MPS we can associate a MPS transfer matrix 
\begin{align}
	\mathcal{T}_{A} = \sum_{i} A^{i} \otimes \bar{A}^{i} = \vcenter{\hbox{
\includegraphics[width=0.12\linewidth]{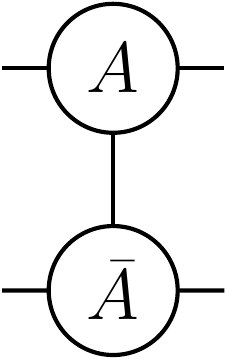}}}, \label{eq:mpstm}
\end{align}
which is a key object in numerical simulations and is used to calculate static correlation functions with respect to a uniform MPS ground state. We henceforth assume that the MPS is normalized such that the transfer matrix has a unique largest eigenvalue equal to 1.

As is well known, a different perspective on the MPS transfer matrix can be provided in terms of a two-dimensional tensor network associated to the uniform MPS representation of the ground state of a local one-dimensional translation invariant Hamiltonian \cite{trotter1976operators,suzuki1976mapping,betsuyaka1984tm,sirker2005path}. For clarity, an overview of this construction is given in \Fig{fig:pathintegral}, where we consider the imaginary time evolution
\begin{align}
	\ket{\Psi_{0}}=\lim_{\beta \to \infty} \frac{\ec^{-\beta H}\ket{\phi}}{||\ec^{-\beta H} \ket{\phi}||}, \label{eq:imagtimeevolv}
\end{align}
where $\ket{\phi}$ is an initial state assumed to have non-zero overlap with the ground state $\ket{\Psi_0}$. Due to the locality of $H$, we can split $\beta$ into small imaginary time steps $\delta$ and use a Trotter-Suzuki decomposition $\ec^{-\delta H} \approx \prod_{n} \ec^{-\delta h_{n,n+1}}$ to arrive at a translation invariant MPO representation of $\ec^{-\delta H}$ (see, e.g., Ref.~\onlinecite{pirvu2010matrix}). This approximation introduces a controllable Trotter error depending on the order of the decomposition, which renders the MPO representation quasi-exact. By grouping tensor contractions along imaginary time, we can identify the {\it exact} MPS transfer matrix with a single column of the network. Interpreting the tensor network in \Fig{fig:pathintegral} as an Euclidean path integral, the exact MPS transfer matrix thus coincides with the quantum transfer matrix derived from the partition function $Z_{\infty} = \lim_{\beta \to \infty} \tr \left( \ec^{-\beta H} \right)$ at zero temperature. Note however that both these exact transfer matrices have exponentially diverging bond dimensions, and thus differ from the truncated finite-dimensional MPS transfer matrix $\mathcal{T}_{\tilde{A}} = \sum_{i} \tilde{A}^{i} \otimes \bar{\tilde{A}}^{i}$ defined before, which arises in actual numerical simulations. Similarly, the exact MPS ground state tensor $A^{i}$ corresponds to a semi-infinite MPO in this picture, and represents the ground state $\ket{\Psi_{0}}$ up to some Trotter error, of which the truncated finite-dimensional MPS $\tilde{A}^{i}$ is a compressed version. Stated in these terms, our goal is to shed light on the relationship between the exact MPS $A^{i}$ and its compressed version $\tilde{A}^{i}$.

\section{Coarse-graining transfer matrices} As it is our intention to coarse-grain a translation invariant MPO, defined by a local tensor $O$ having on-site operator dimension $d$ and MPO bond dimension $D$, a natural way to proceed is to devise a coarse-graining strategy using MPOs in order to retain the matrix product structure. To this end, we introduce an isometric coarse-graining MPO denoted by $G$, which is characterized locally by a single five-index tensor $g$,

\begin{align}
\vcenter{\hbox{
\includegraphics[width=0.85\linewidth]{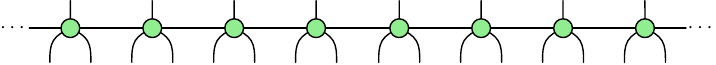}}} \label{eq:mpoisometry}
\end{align}

The tensor $g$ is restricted to be an isometry defined by the map $g : \mathbb{I} \otimes \mathbb{I} \otimes \mathbb{V} \to \mathbb{O} \otimes \mathbb{V}$, where $\mathbb{I}$, $\mathbb{V}$, and $\mathbb{O}$, respectively, refer to the vector spaces of the incoming indices, the virtual indices and the outgoing indices. As such, $g$ satisfies $g{^\dagger} g = \openone_{\mathbb{O} \otimes \mathbb{V}}$ and $g g{^\dagger} = P$, where $P$ is a projector onto some subspace of $\mathbb{I} \otimes \mathbb{I} \otimes \mathbb{V}$. Let us denote $d=\dim (\mathbb{I})$, $\chi=\dim (\mathbb{V})$ and $d'=\dim (\mathbb{O})$. The isometric constraints on the local tensor $g$ forces the MPO $G$ as a whole to be isometric if $d^{2} > d'$, or unitary if $d^{2} = d'$. Note that there are two different ways to group operator and virtual indices of $g$, leading to two possible ``gauge'' choices, denoted pictorially by the following equations:

\begin{align}
\vcenter{\hbox{
\includegraphics[width=0.30\linewidth]{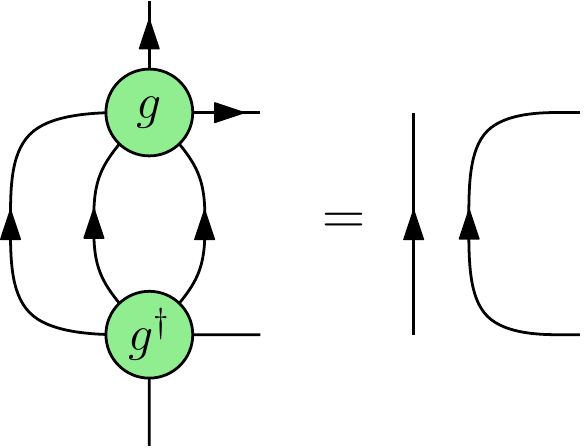}}} \mathrm{\quad or \quad} \vcenter{\hbox{
\includegraphics[width=0.30\linewidth]{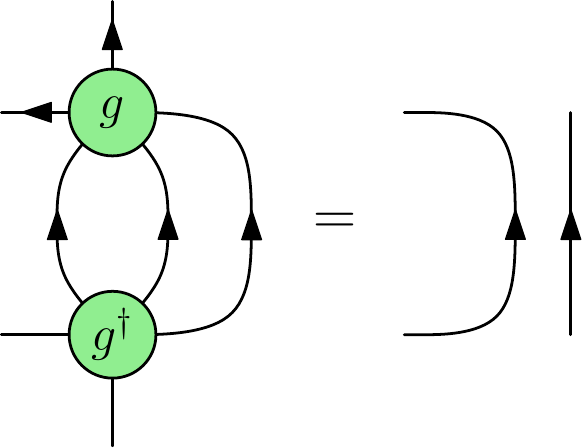}}}\quad . \label{eq:isometries}
\end{align}

Let us now consider a finite set $\{G_{s}\}$ of the kind of MPO isometries defined above, and label them with a discrete scale dimension $s = 1,2, \ldots, s_{\rm max}$, where each $G_{s}$ is allowed to be different. By acting sequentially with each of the $\{G_{s}\}$ together with their conjugates $\{G_{s}^{\dagger}\}$ on the translation invariant MPO to be coarse-grained, we arrive at the MERA-inspired tensor network (see Appendix~\ref{app:mera}):

\begin{align}
\vcenter{\hbox{
\includegraphics[width=0.8\linewidth]{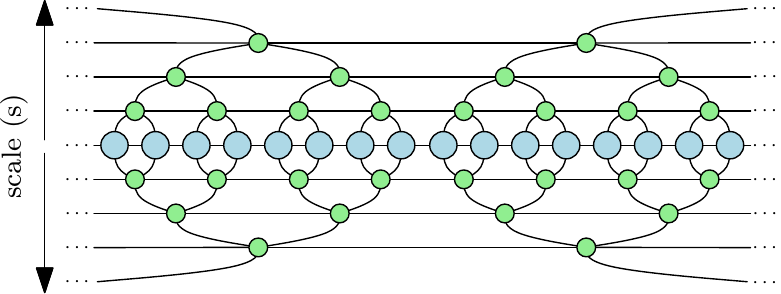}}} \label{eq:tmansatz}
\end{align}

One way to optimize all of these coarse-graining MPOs, is to perform a single sweep in scale from bottom to top. At each layer $s$, two tensors $O_{s-1}$ of the previous layer are blocked by acting with $g_{s}$ and $g_{s}^{\dagger}$ to construct a coarse-grained tensor
\begin{align}
\vcenter{\hbox{
\includegraphics[width=0.16\linewidth]{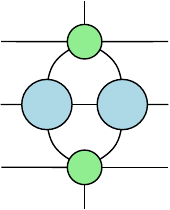}}} \label{eq:tmobject},
\end{align}
which redistributes entanglement among virtual and operator dimensions. Tracing over the outgoing indices, as depicted in \Eq{eq:costfunction}, we can interpret this object as a generalized transfer matrix with $\chi_{s}^{2}D_{s}$-dimensional fixed points $(\sigma_{L}[g_{s}]|$ and $|\sigma_{R}[g_{s}])$, where we have emphasized the dependence on the coarse-graining isometry in the notation. By locally maximizing\footnote{Like all tensor network optimization problems, there is no guarantee that a global optimal solution will be found. For our purposes, the optimization procedure was seen to consistently converge to a local optimum.} the cost function \footnote{This cost function is well-defined for Hermitian MPOs; a possible alternative strategy for non-Hermitian MPOs would be to minimize the norm of the difference between two tensors $O_{s-1}$ and the same two tensors with the projector $P_{s}=g_{s} g_{s}^{\dagger}$ applied to the outgoing indices.}
\begin{align}
C(g_{s})=\vcenter{\hbox{
\includegraphics[width=0.25\linewidth]{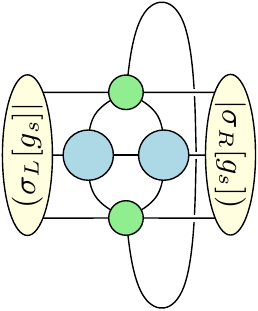}}} \label{eq:costfunction}
\end{align}
for the isometry $g_{s}$ using a conjugate gradient algorithm adapted to unitary manifolds \cite{abrudan2009conjugate}, we then iteratively update the left and right fixed points of \Eq{eq:tmobject} until convergence of \Eq{eq:costfunction} is attained up to some tolerance. The coarse-grained MPO tensor $O_{s}$ is then obtained by truncating its bond dimension to a fixed number $D_{s}$, or up to some tolerance, using conventional MPS methods \cite{orus2008itebd}, such that
\begin{align}
\vcenter{\hbox{
\includegraphics[width=0.60\linewidth]{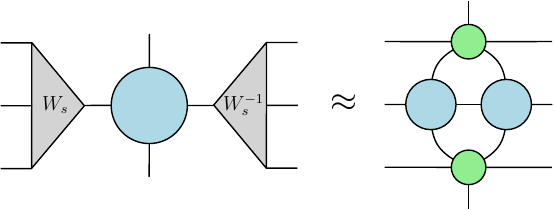}}},
\end{align}
where we have introduced the rank-reducing tensor $W_{s}$, with $W_{s}^{-1}$ denoting the left inverse of $W_{s}$ such that $W_{s}^{-1}W_{s}=\openone$ and $W_{s}W_{s}^{-1}$ projects onto the truncated subspace. Note that the new on-site operator dimension of the coarse-grained MPO is determined by the output dimension $d'_{s}$ of the gate $g_{s}$. This constitutes the optimization of one layer, and by repeating the above coarse-graining procedure until the top level is reached, we obtain a set of effective MPO tensors $\{O_{s}\}$, isometric gates $\{g_{s}\}$ and truncation tensors $\{W_{s}\}$ for $s=1,2,\ldots,s_{\rm max}$. Both the tolerances for the optimization of the cost function and for the truncations provide sensible measures for the errors introduced along the way.

Although our main motivation in this article is to coarse-grain the MPOs appearing naturally along the imaginary time direction in the Euclidean path integral picture in \Fig{fig:pathintegral}, there is nothing preventing us from applying the above ansatz along the spatial direction to renormalize, for instance, a Hamiltonian operator or more general transfer matrices arising in two-dimensional lattice models. In particular, local Hamiltonian terms can be seen to renormalize to a sequence of semi-infinite MPO strings to the left or to the right depending on the gauge choice in \Eq{eq:isometries}, which follows naturally from the one-sided causal cone structure that arises due to the isometric nature of the coarse-graining gates.

\begin{figure*}[t]
 \centering
 \includegraphics[width=1.0\linewidth,keepaspectratio=true]{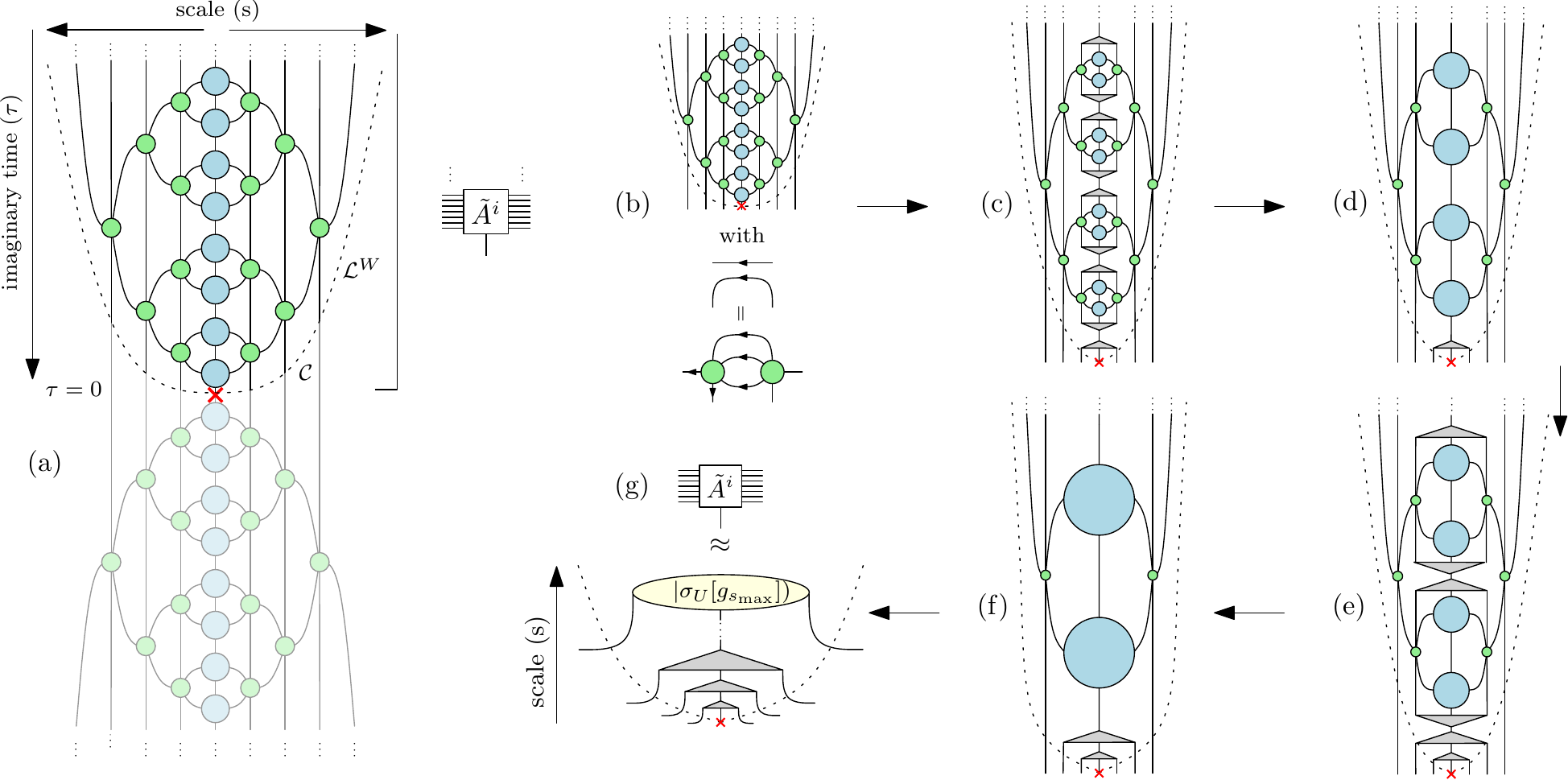}
 \caption{(Color online) Uniform MPS from coarse-graining the exact transfer matrix. Location of the physical spin, considered as an impurity at $\tau=0$, is denoted with a red cross. (a) Bonds crossing the one-sided causal cone $\mathcal{C}$ of the impurity at $\tau=0$ identify degrees of freedom relevant to the physical spin, which live on the Wilson chain $\mathcal{L}^{W}$ defined along the boundary of $\mathcal{C}$. The tensor network outside $\mathcal{C}$ renormalizes the exact transfer matrix $\mathcal{T}_{A}$ into a new transfer matrix $\mathcal{T}_{W}$ along $\mathcal{L}^{W}$. (b) Contraction of the tensor network outside $\mathcal{C}$ renormalizes the ket, projecting the exact MPS $A^{i}$ onto the subspace defined by the relevant degrees of freedom along $\mathcal{L}^{W}$. (c) Insertion of approximate resolutions of the identity $\openone \approx W_{1}W_{1}^{-1}$. (d) Coarse-graining the semi-infinite MPO string to the next layer leaves a single $W_{1}$ tensor behind. (e)(f) Repeating this procedure leads to the construction of the Wilson MPO. (g) After introducing an infrared cutoff by capping the Wilson MPO with the fixed point of the uppermost coarse-grained tensor, we retrieve a finite-dimensional MPS $\tilde{A}^{i}$.}
 \label{fig:ansatz}
\end{figure*}

\section{From Euclidean path integral to uniform MPS} \label{sec:path2umps}
The pioneering work of Wilson on the numerical renormalization group (NRG) showed that the relevant low-energy subspace for an impurity problem could be obtained by applying real-space RG transformations \cite{wilson1975nrg}. Recently, the theory of minimal updates in MERA has related the success of NRG to the inherent causal cone structures arising in MERA \cite{evenbly2015theory}. The causal cone of a region had been originally introduced as the part of the MERA network that is geometrically connected to and able to exert influence on the properties of the state in that region \cite{vidal2007entanglement}. It has since been interpreted as the collection of tensors that needs to be changed in order to account for a local change of the Hamiltonian in that region, which is understood to be sufficient to capture the evolution of that region under successive coarse-graining transformations while maintaining locality \cite{evenbly2015theory}. This property can be ultimately traced back to the existence of distinct energy scales in the Hamiltonian. According to the principle of minimal updates, an impurity initially localized in space thus remains localized under coarse-graining, which leads to a very efficient MERA description of systems with boundaries, impurities, or interfaces \cite{evenbly:2010boundary,evenbly2014impurities}. The relevant degrees of freedom for an impurity are found to be exactly those living at the boundary of the causal cone.

Although the MPS ansatz we are about to describe bears a resemblance to the impurity MERA construction, the underlying motivation is entirely different. We stress that the main motivation for our proposal is to be able to capture the physics relevant for the physical spin degree of freedom of a quantum state, which, as we will argue, precisely amounts to extracting the degrees of freedom relevant to the spin treated as an impurity in the exact quantum transfer matrix.

\subsection{Construction of MPS ansatz} \label{subsec:constransatz}
Let us now apply our coarse-graining ansatz \Eq{eq:tmansatz} to the vertical transfer matrix $\mathcal{T}_{A}$ obtained from the imaginary time evolution network in \Fig{fig:pathintegral}. Note that the physical dimension and virtual dimension in \Fig{fig:pathintegral} now correspond, respectively, to the bond dimension and the operator dimension of the coarse-graining MPO. We furthermore implicitly assume that $\mathcal{T}_{A}=\ec^{-\tilde{H}}$, i.e.~the transfer matrix is understood to originate from some Euclidean rotated effective local Hamiltonian $\tilde{H}$ \cite{zauner2015transfer}. In general, there is no reason to expect the Hamiltonian $\tilde{H}$ to be related to the physical Hamiltonian $H$ involved in the MPO description of the imaginary time evolution $\ec^{-\delta H}$.

To arrive at a uniform MPS, we will treat the physical spin connecting the exact MPS representations of ket and bra at $\tau=0$ as an impurity in $\mathcal{T}_{A}$. The arbitrary location of the physical index in imaginary time does not \textit{a priori} introduce an inhomogeneity in $\mathcal{T}_{A}$. However, for expectation values of local operators different from the identity, the privileged nature of the physical spin becomes manifest, and an operator insertion at $\tau=0$ leads to a modified transfer matrix $\mathcal{T}_{A}^{\mathcal{O}} = \sum_{ij} \mathcal{O}_{ij} A^{i} \otimes \bar{A}^{j}$ used in calculating MPS expectation values. As such, a static correlation function between two operators separated by $n$ sites in the physical system corresponds to a ``temporal correlator'' of the impurity between two operators separated by $n$ steps of evolution with $\mathcal{T}_{A}=\ec^{-\tilde{H}}$, where $\tilde{H}$ again denotes not the physical Hamiltonian $H$ but the Euclidean rotated effective local Hamiltonian. We will thus construct a truncated MPS representation $\tilde{A}^i$ by extracting the relevant degrees of freedom for the ``Euclidean dynamics'' of the impurity from the exact quantum transfer matrix $\mathcal{T}_{A}$.

In what follows, we will focus on the upper semi-infinite MPO describing the ket part $\ket{\Psi[A]}$ of the transfer matrix $\mathcal{T}_{A}$, as depicted in \Fig{fig:ansatz}, and refer to the location of the physical spin as the ``impurity" regardless of whether there is actually an operator inserted at $\tau=0$. To explicitly construct the approximated ket tensor $\tilde{A}^{i}$ associated to the uniform MPS $\ket{\Psi [\tilde{A}]}$ in \Fig{fig:pathintegral}, we first apply our ansatz to the infinite quantum transfer matrix, and optimize for $s_{\rm max}$ layers as if no impurity were present. We then insert the impurity at $\tau=0$ and draw its causal cone $\mathcal{C}$ in \Fig{fig:ansatz}(a), where the inside of the causal cone stretches out to the left and to the right, and the degrees of freedom affecting the impurity at $\tau=0$ arise from contracting the network outside the causal cone. By ignoring the bottom half of the network, we can identify the part of the coarse-graining network which acts on the semi-infinite ket part of the MPO. In retrospect, the impurity naturally suggests which isometric restriction in \Eq{eq:isometries} to impose on the coarse-graining tensors during the optimization, as the entire tensor network surrounding the original semi-infinite transfer matrix can be interpreted as a projector onto the dominant eigenvector subspace relevant to the impurity, see \Fig{fig:ansatz}(b).

It is clear that only one index of the MPO coarse-graining tensors crosses the boundary of the causal cone for each layer. Together, these legs constitute the sites of an effective lattice system $\mathcal{L}^{W}$ defined along the boundary of the causal cone, which is called the Wilson chain. Sites along this chain are labeled by the layer index $s=1,2,\ldots,s_{\rm max}$, where site $s$ contains an effective renormalized description, for $d'=d$, of $2^{s}$ sites of the original lattice located roughly at a distance $2^{s}$ away from the impurity. Moving along the Wilson chain thus corresponds to changing scale and moving away from the impurity. Next, we insert approximate resolutions of the identity at the lowest layer in \Fig{fig:ansatz}(c) using the rank-reducing tensors $W_{1}W_{1}^{-1}$ obtained during the optimization to reduce the bond dimension in the imaginary time direction. As shown in \Fig{fig:ansatz}(d), the semi-infinite MPO string is pushed to the next coarse-graining layer, leaving a single tensor $W_{1}$ behind. Repeating this procedure for the next layer in \Fig{fig:ansatz}(e)-(f) leads to the emergence of an inhomogeneous MPO along the imaginary time direction, which we will refer to as the {\it Wilson MPO}, and which effectively amounts to a sequence of coarse-grained Trotter steps \footnote{Even though the bond dimension $\prod_{s=1}^{s_{\rm max}} \chi_{s}$ of $\tilde{A}^{i}$ may become large for large $s_{\rm max}$, we can efficiently contract the Wilson MPO sequentially and accurately truncate its bond dimension to some value $D$. We could partly avoid this additional truncation, which is a consequence of having MPOs corresponding to infinitesimal Trotter times $\delta$, by blocking MPOs initially such that they represent a bigger time step \cite{evenblyprivate}.}.

By introducing the fixed point of the uppermost coarse-grained tensor as an infrared cutoff in \Fig{fig:ansatz}(g), we arrive at a finite-dimensional approximate Wilson-based MPS with an internal layered structure resulting in the uniform MPS tensor $\tilde{A}^{i}$. The reason for this particular cutoff strategy can be traced back to interpreting the transfer matrix $\mathcal{T}_{A}$ as a thermal state (or more generally a mixed state) with exponentially decaying correlations. We then expect the coarse graining network to be able to disentangle this state into a product state using a finite number of layers, similar as in the case of ground states of gapped Hamiltonians, such that, after a finite number of layers, we can pictorially denote the flow by
\begin{align}
\vcenter{\hbox{
\includegraphics[width=0.40\linewidth]{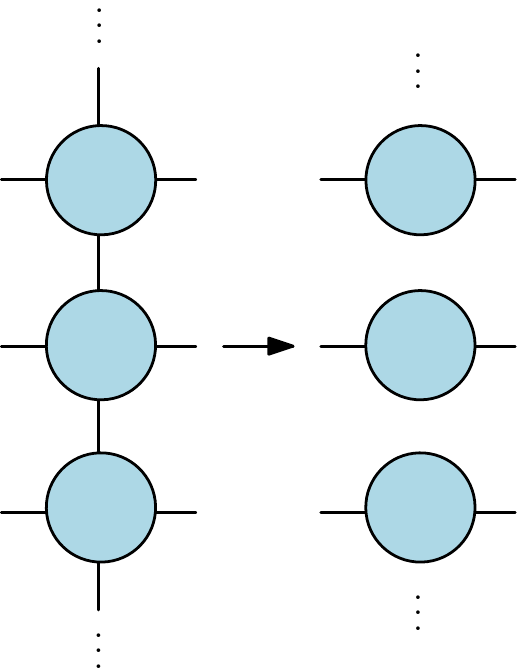}}}
\end{align}
If we want the coarse-grained MPO tensor at the top to have bond dimension $D_{s_{\rm max}}=1$, the optimal choice is given by the eigenvectors corresponding to the largest eigenvalue of \Eq{eq:tmobject}. For a critical MPO, we expect scale invariance to direct the coarse graining process to a fixed point that needs to be iterated forever. In that case, introducing the product state fixed point as an infrared cutoff explicitly breaks the critical properties and can as such only be considered an approximation, whose quality can be assessed by evaluating the energy with respect to the resulting state. Note that all variational, finite bond dimension MPS approximations necessarily contain an implicit infrared cutoff, and that different cutoff implementations might yield different MPS tensors approximating the same gapless state, a point to which we return in \Sec{sec:mpoansatz}.

In the following two sections, we consider two immediate applications made possible by the structure of our ansatz.

\subsection{Structure of MPS fixed point reduced density matrices} \label{sec:fp}
From the layered structure of our MPS decomposition, it is straightforward to associate a layered decomposition to the zero-dimensional fixed point reduced density matrices $\rho_{L}$ and $\rho_{R}$ of the MPS transfer matrix \Eq{eq:mpstm} as well. To see this, let us apply a truncation procedure to the transfer matrix $\mathcal{T}_{\tilde{A}}$ constructed from our ansatz, instead of working solely on the level of the ket $\ket{\Psi[\tilde{A}]}$. By sequentially grouping indices across bra and ket for the truncation step at each layer, we eventually arrive at the norm
\begin{align}
\braket{\Psi[\tilde{A}]|\Psi[\tilde{A}]}=\vcenter{\hbox{
\includegraphics[width=0.60\linewidth]{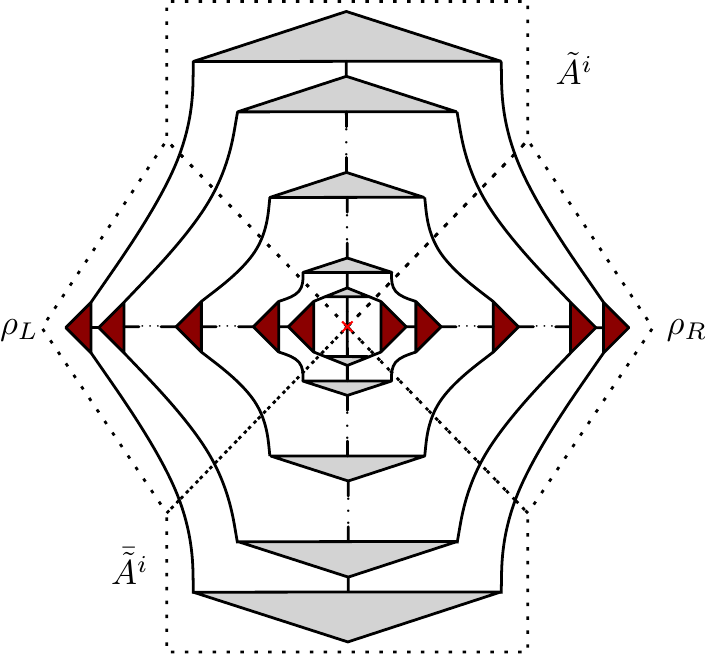}}}, \label{eq:fpdecomposition}
\end{align}
where the grey triangles denote the Wilson MPOs for ket and bra, and the red triangles 
are now invertible rank-reducing tensors $X^{-1} _{s}$ (left) and $X_{s}$ (right), for $s=1,2,\ldots,s_{\rm max}$, that can be obtained from lifting the impurity from bottom to top and truncating the effective transfer matrix at each layer $s$ by considering it as a MPS, 
\begin{align}
\vcenter{\hbox{
\includegraphics[width=0.80\linewidth]{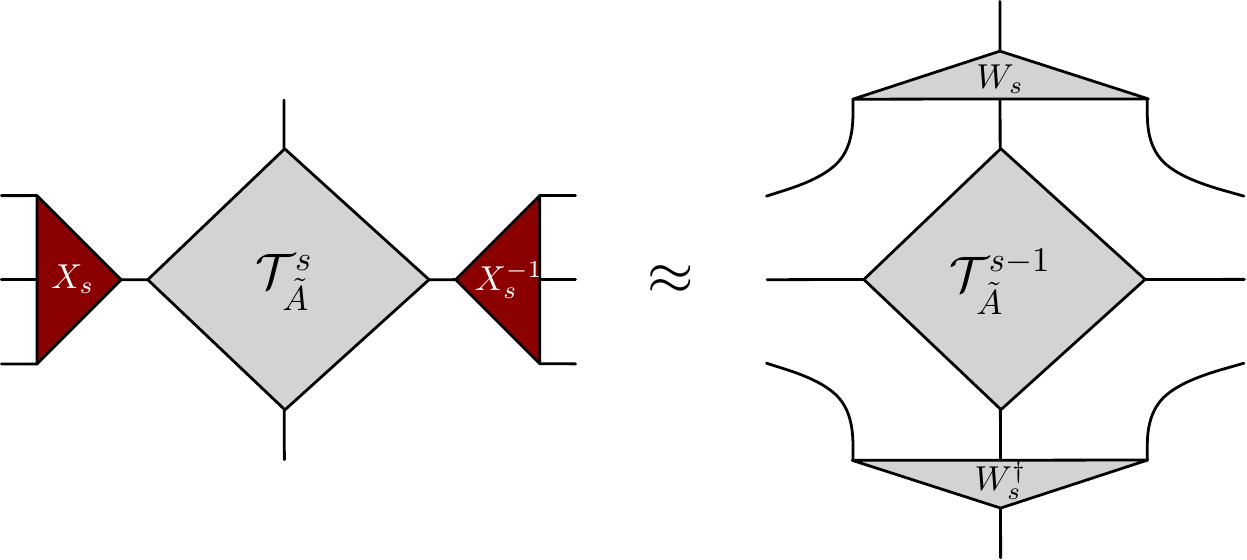}}}. \label{eq:superoperator}
\end{align}
Note how we can approximately determine the fixed points of the transfer matrix in a sequential manner without explicitly calculating them, and how only small bond dimensions are involved in the intermediate truncation processes. A potential application of this insight lies in the problem of efficiently contracting two-dimensional tensor networks, \textit{e.g.}~overlaps of projected entangled pair states \cite{verstraete2004renormalization}, where the one of the main bottlenecks in variational ground state optimizations can be traced back to calculating high-dimensional MPS fixed points of the MPO transfer matrices needed for the calculation of environment tensors. The feedback of information from high- to low-energy in our layered decomposition of these fixed points might lead to a substantial reduction in computational resources, thus allowing the approximation of otherwise intractable MPS fixed points \cite{bal2015inprepuniversality}. 

Another application concerns scale invariant theories, for which we expect to find a recursive relation, directly from the layered MPS decomposition, whose eigendecomposition ought to contain the scaling operators and scaling dimensions of the critical theory that we are approximating. Suppose we have identified a fixed point truncation tensor $W^{*}$ and associated $X^{*}$ in the layered decomposition of our MPS ansatz, characterizing the scale invariant fixed-point behavior. We can then propose a ``radial'' superoperator
\begin{align}
\vcenter{\hbox{
\includegraphics[width=0.40\linewidth]{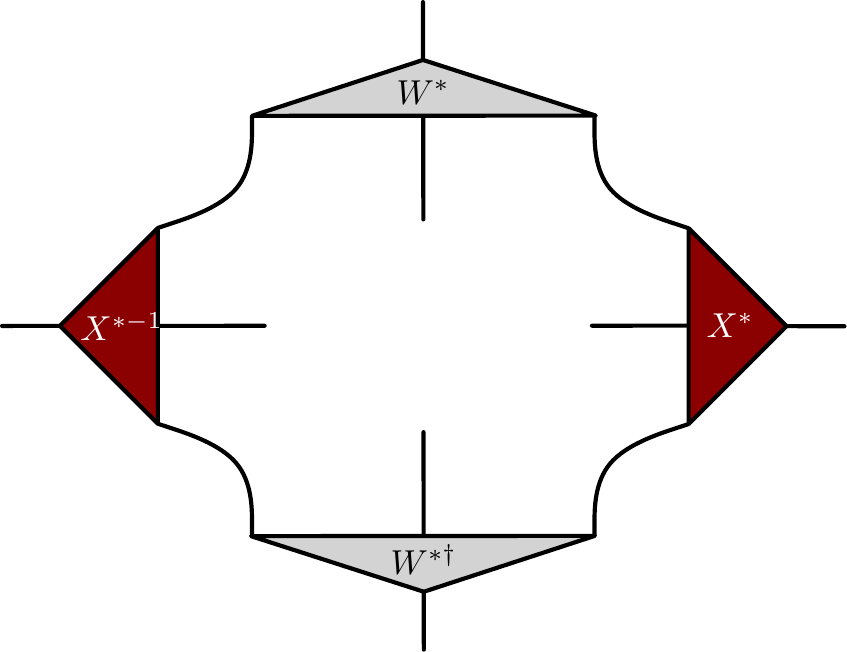}}}, \label{eq:scalefp}
\end{align}
whose eigenvalues and eigenvectors are expected to be related to the scaling dimensions and operators of the underlying conformal field theory. Forgetting about the Wilson-based MPS origin of the ansatz, we can also build concentric tensor networks from \Eq{eq:scalefp}, which enforce scale invariance and can be regarded as variational ans\"atze in their own right \cite{bal2015inprepuniversality,huebener2010concatenated}. Similar superoperators for extracting scaling fields were of course conceived in MERA \cite{giovannetti2008quantum,pfeifer2009entanglement} and TNR \cite{evenbly2014tnr}, but not in a MPS setting. A partial approach to extract scaling information from a MPS was investigated in Refs.~\onlinecite{tagliacozzo2008scaling,pollmann2009theory,lauchli2013operator,stojevic2015conformal}.

\subsection{Restricted variational subspaces for excitations} \label{subsec:resvarsubex}
Given the internal layered structure of the uniform MPS in \Fig{fig:ansatz}(g), it is tempting to ask how excitations can be interpreted in this framework. Following Wilson's RG interpretation of the Kondo impurity problem \cite{wilson1975nrg}, we expect low-energy excitations to live near the top of the Wilson MPO if the layered MPS decomposition obtained from our ansatz is to be interpreted as a true renormalization group scale. Note that our ansatz retains translation invariance in space as we effectively construct a uniform MPS, and that the scale dimension in our case refers to coarse-grained imaginary time. We want to emphasize that, even though there is no explicit spatial coarse-graining taking place (as the number of sites remains constant), spatial correlations do arise naturally in our picture as we coarse-grain imaginary time to grow the virtual bond dimension of the MPS. Despite the absence of an explicit coarse-graining in space, the successive MPO layers labeled by $s$ are non-unitary and can therefore shrink the effective Hilbert space at every step. The idea is now to expose this effective reduction of the Hilbert space at higher renormalization scales by perturbing the tensors within the layered structure of the MPS and observing to which part of the spectrum of the Hamiltonian this particular perturbation gives access.

In particular, exploiting the translation invariance which we still have at our disposal, we can use the variational MPS ansatz for localized excitations developed in Refs.~\onlinecite{haegeman2012variational,haegeman2013postmps}, given by
\begin{align}
	&\quad \ket{\Phi^{(A)}_{p}(B)} = \nonumber \\  &\sum_{n \in \mathbb{Z}}^{} \ec^{\ic p n} \sum_{{i_n}=1}^{d} \mathbf{v}^{\dagger}_{L} \left( \prod_{m < n} A^{i_{m}} \right) B^{i_n} \left( \prod_{m' > n} A^{i_{m'}} \right) \mathbf{v}_{R} \ket{\mathbf{i}}. \label{eq:mpsexc}
\end{align}
Comparing with \Eq{eq:mps}, it is clear that the excitation ansatz is constructed on top of the ground state wavefunction, as we change a single ground state tensor $A^{i}$ into a tensor $B^{i}$ and take a momentum superposition of this localized perturbation. All variational freedom is contained within the tensor $B^{i}$, and the variational optimization of the Rayleigh-Ritz quotient
\begin{align}
	\min_{B} \frac{\braket{\Phi^{(A)}_{p}(B)|H|\Phi^{(A)}_{p}(B)}}{\braket{\Phi^{(A)}_{p}(B)|\Phi^{(A)}_{p}(B)}}. \label{eq:geig}
\end{align}
gives rise to a generalize eigenvalue problem. In order for this generalized eigenvalue problem to be well-defined, it will be necessary to project out so-called null-modes, i.e.\ (almost) zero eigenvalues of the effective normalization matrix in the right hand side of the generalized eigenvalue equation.

\begin{figure}[t]
 \centering
 \includegraphics[width=1.0\linewidth,keepaspectratio=true]{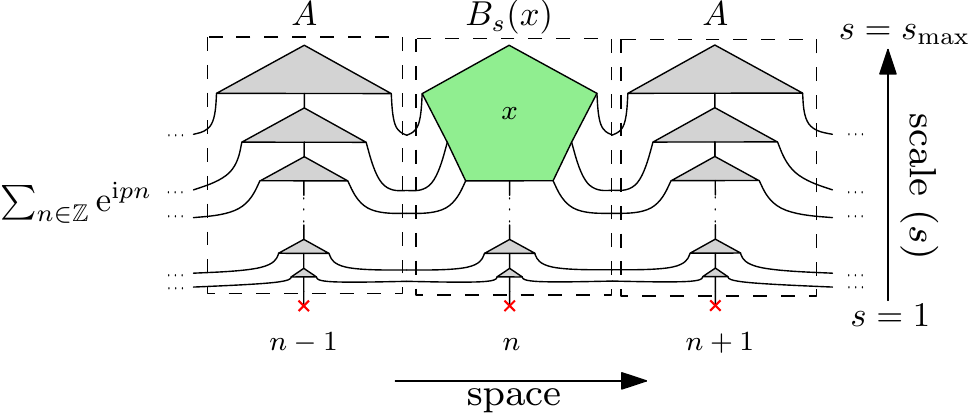}
 \caption{(Color online) Example of a restricted variational MPS excitation ansatz $\ket{\Phi^{(A)}_{p}(B_{s})}$ at intermediate layer index $s$, constructed to allow for targeting the variational parameters $x$ residing in an interval of layers $[s,s_{\rm max}]$ (green triangles), while the tensors corresponding to layers $[1,s)$ are held fixed.}
 \label{fig:fullexcansatz}
\end{figure}

Let us now recast \Eq{eq:mpsexc} pictorially and substitute our layered MPS decomposition of the MPS ground state to arrive at the variational ansatz depicted in \Fig{fig:fullexcansatz}. By fixing part of $B^{i}$ to be equal to $A^{i}$ using the layered decomposition, we can now design a restricted excitation ansatz $\ket{\Phi^{(A)}_{p}(B_{s})}$, for $s=1,2,\ldots,s_{\rm max}$, which only allows for variations with respect to an interval $[s,s_{\rm max}]$ of tensors in the Wilson MPO description of the full excitation ansatz. The restricted tensor $B^{i}_{s}(x) \equiv \Xi_{s} (x)$ is defined in terms of a linear mapping
\begin{align}
\Xi_{s} : \mathbb{C}^{\prod_{s'=s}^{s_{\rm max}} \chi_{s'} \times d_{s} \times \prod_{s'=s}^{s_{\rm max}} \chi_{s'}} \to \mathbb{C}^{D \times d \times D},
\end{align}
which glues the variational parameters $x$ to the surrounding MPS ground state tensors $A^{i} \in \mathbb{C}^{D \times d \times D}$. For different layers $s$, this tensor parametrizes different subspaces of the full variational space\footnote{Note that the restricted generalized eigenvalue problem is variational with respect to using the full $B^{i}$, but not necessarily with respect to the exact problem. Although the reduced $B^{i}(x)$ will always yield excitation energies higher than those obtained by varying the full $B^{i}$, these energies might still be lower than the exact excitation energies due to errors in the ground state MPS approximation \cite{haegeman2013postmps}} spanned by the states $\ket{\Phi^{(A)}_{p}(B)}$ with unrestricted $B^{i}$. Starting from $B^{i}_{s_{\rm max}}$, where only the variational degrees of freedom in the top tensor are considered, we can gradually take all layers $[1,\ldots,s_{\rm max}]$ into account until the ansatz culminates into the full $B^{i} \equiv B^{i}_{1}$.

Note that the role of our restricted ansatz is not so much in improving the efficiency of numerically calculating excitations within the framework established in Ref.~\onlinecite{haegeman2013postmps}, but in providing a novel and conceptually intriguing interpretation of excitations within MPS in a way that explicitly tries to resolve the different energy scales present in the MPS ground state tensor. In \Sec{sec:resvarsubex}, we will verify that the number of null-modes as function of layer index $s$ and momentum $p$ is intimately tied to the effective reduction of the Hilbert space at higher renormalization scales.

\section{Numerical Results}

To assess the validity of our ansatz, we have performed numerical simulations using the MPS ansatz introduced in \Sec{subsec:constransatz} and discuss the transfer matrix and Schmidt spectra of the resulting MPS ground states. We calculate dispersion relations using the setup of \Sec{subsec:resvarsubex} to illustrate that variational subspaces corresponding to the upper layers of our MPS ansatz are effectively restricted to the low-energy part of the spectrum. Additionally, we translate our method for compressing transfer matrices to the setting of free fermions, where the additionally imposed structure on the numerics allows us to highlight and further corroborate our findings.

\subsection{MPS ansatz from Wilson MPO} \label{sec:mpoansatz}
\begin{figure}[t]
 \centering
 \includegraphics[width=\linewidth,keepaspectratio=true]{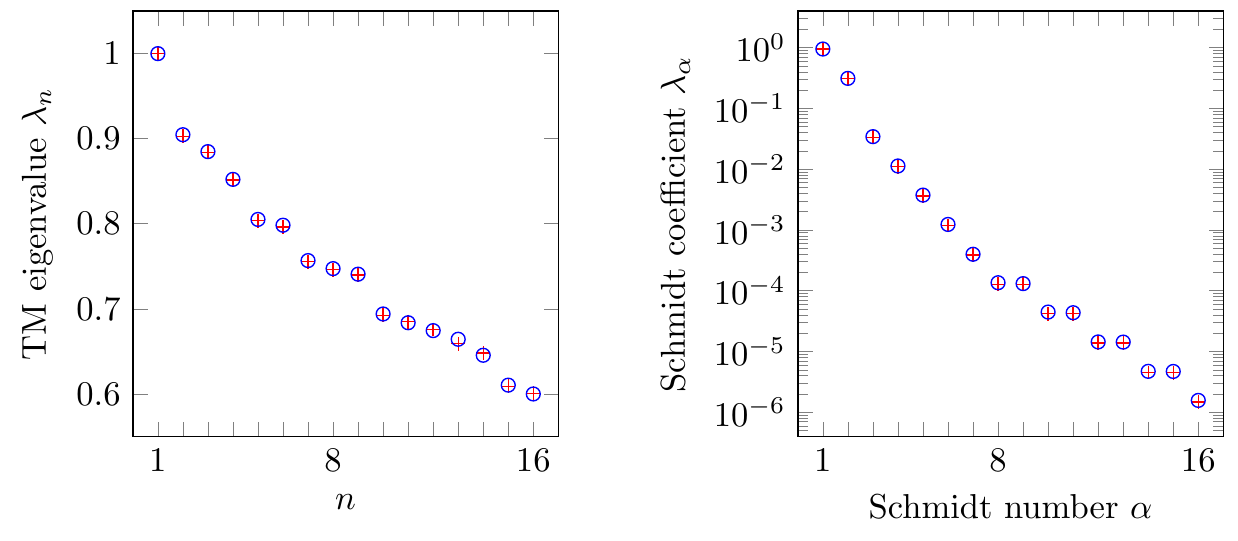}
 \caption{(Color online) Comparison of transfer matrix (TM) spectra (left) and Schmidt spectra (right) of the transverse Ising model \Eq{eq:ising} between our MPS ansatz $\ket{\Psi[\tilde{A}]}$ based on the Wilson MPO (blue circles) and a variational MPS ground state $\ket{\Psi[A]}$ (red crosses) for $\lambda=1.1$ (gapped) and $D=16$. All MPO simulations were performed using a Trotter step $\delta=0.001$ and local dimensions $d_s, \chi_s, d'_s \in \{ 2,4 \}$ of the coarse-graining tensors with $s_{\rm max}=14$ layers.}
 \label{fig:spectragapped}
\end{figure}

\begin{figure}[t]
 \centering
 \includegraphics[width=\linewidth,keepaspectratio=true]{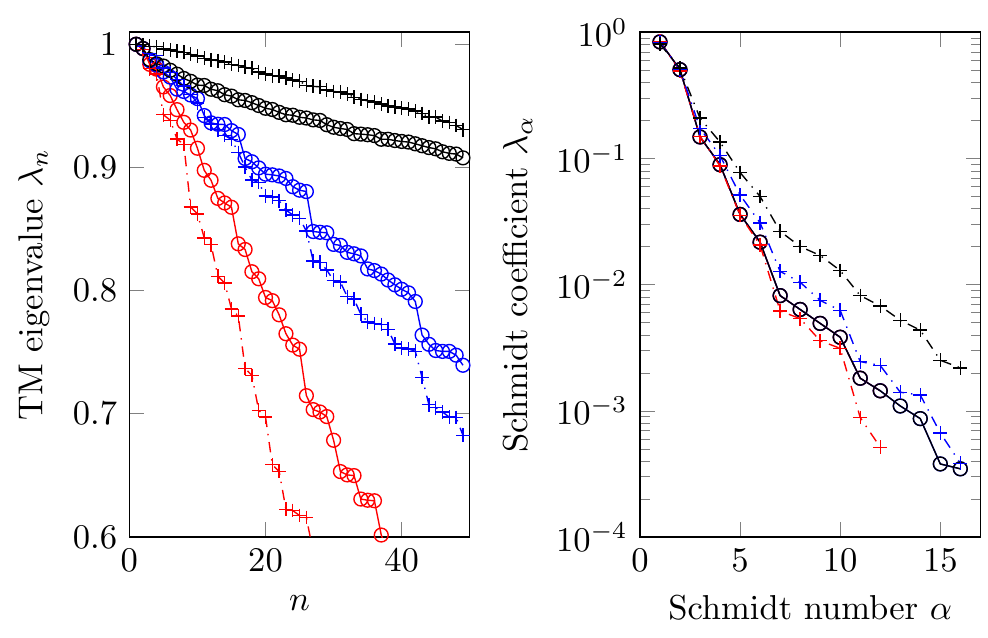}
 \caption{(Color online) Comparison of transfer matrix (TM) spectra and Schmidt spectra of the transverse Ising model \Eq{eq:ising} between our MPS ansatz $\ket{\Psi[\tilde{A}]}$ based on the Wilson MPO (black circles) and a variational MPS ground state $\ket{\Psi[A]}$ (crosses) with bond dimension $D=12$ (red), $D=24$ (blue) and $D=50$ (black) for $\lambda=1$ (gapless). All MPO simulations were performed using a Trotter step $\delta=0.001$ and local dimensions $d_s, \chi_s, d'_s \in \{ 2,4 \}$ of the coarse-graining tensors with $s_{\rm max}=17$ layers, and we also investigated the effect of further truncating the MPO based ansatz to specific bond dimensions $D=12$ (red circles) and $D=24$ (blue circles) based on the Schmidt coefficients.}
 \label{fig:spectragapless}
\end{figure}

Consider the quantum Ising model in a transverse field, which can be defined in terms of the Hamiltonian 
\begin{align}
	H = -\sum_{i} \sigma_{i}^{x} \sigma_{i+1}^{x} - \lambda \sum_{i} \sigma_{i}^{z},\label{eq:ising}
\end{align}
where $\lambda$ determines the strength of the applied magnetic field. One can easily show how to construct a translation invariant MPO representation\cite{pirvu2010matrix} of the Trotter-Suzuki decomposition of $\ec^{-\delta H}$, to which we apply our ansatz along the imaginary time direction. Comparing the ground state energy density of the MPS resulting from our ansatz with the exact solution, we find relative energy errors $\Delta E = (E-E_{0})/E_{0} \sim 10^{-6}$ both for the gapped ($\lambda \neq 1$) and gapless ($\lambda=1$) phases using very modest computational resources. Having obtained a uniform MPS from the Wilson MPO, we can further quantify the accuracy of our ansatz by studying the spectrum of the transfer matrix \Eq{eq:mpstm}, and the Schmidt values of the MPS, which can be retrieved from the spectrum of the dominant eigenvector of the transfer matrix \Eq{eq:mpstm}. In \Fig{fig:spectragapped} and \Fig{fig:spectragapless}, we show a comparison of low-lying transfer matrix spectra obtained with respectively our coarse-graining method and a variationally optimized MPS using the time-dependent variational principle \cite{haegeman2011tdvp}.

For the gapped case depicted in \Fig{fig:spectragapped}, an excellent match is found for both the TM eigenvalues $\lambda_n$ and the Schmidt coefficients $\lambda_{\alpha}$ as soon as the number of coarse-graining layers is chosen sufficiently large, which demonstrates the behavior of the infrared cutoff as explained in \Sec{subsec:constransatz}.

For the gapless case in Fig.~\ref{fig:spectragapless}, the infrared cutoff can be recognized as a very particular initial state in the effective imaginary time evolution induced by the Wilson MPO, potentially spoiling the relevant infrared data for critical models. From the impurity point of view, there is a strong interaction between the different sites in the Wilson chain $\mathcal{L}^{W}$, corresponding to degrees of freedom living at different energy scales. The infrared cutoff introduced after a finite number of layers at one end of the Wilson chain, has a strong feedback on the physics of the impurity, which is living at the other end of the Wilson chain. As argued in \Sec{subsec:constransatz}, different implementations of the implicit infrared cutoff for gapless states, yield different variational MPS approximations, which has recently been discussed in Ref.~\onlinecite{rams2014truncating}.

To illustrate this explicitly, we also studied the effect of further truncating the Wilson based MPS to some specific bond dimensions using the standard MPS recipe (\textit{i.e.}\ throwing away the smallest Schmidt coefficients). While this has no distinguishable effect on the remaining Schmidt coefficients of Fig.~\ref{fig:spectragapless}(b), the effect on the nondominant part of the spectrum of the transfer matrix itself is significant, as can be observed in Fig.~\ref{fig:spectragapless}(a). Nevertheless, when comparing to a variational MPS with a particular bond dimension, we obtain a qualitative agreement that matches the first few dominant Schmidt coefficients (see also \Fig{fig:EntSpectrum_ff}b for analogous entanglement spectra obtained for critical free fermions).

\subsection{MPS excitation ansatz from Wilson MPO} \label{sec:resvarsubex}

\begin{figure}[t]
 \centering
 \includegraphics[width=\linewidth,keepaspectratio=true]{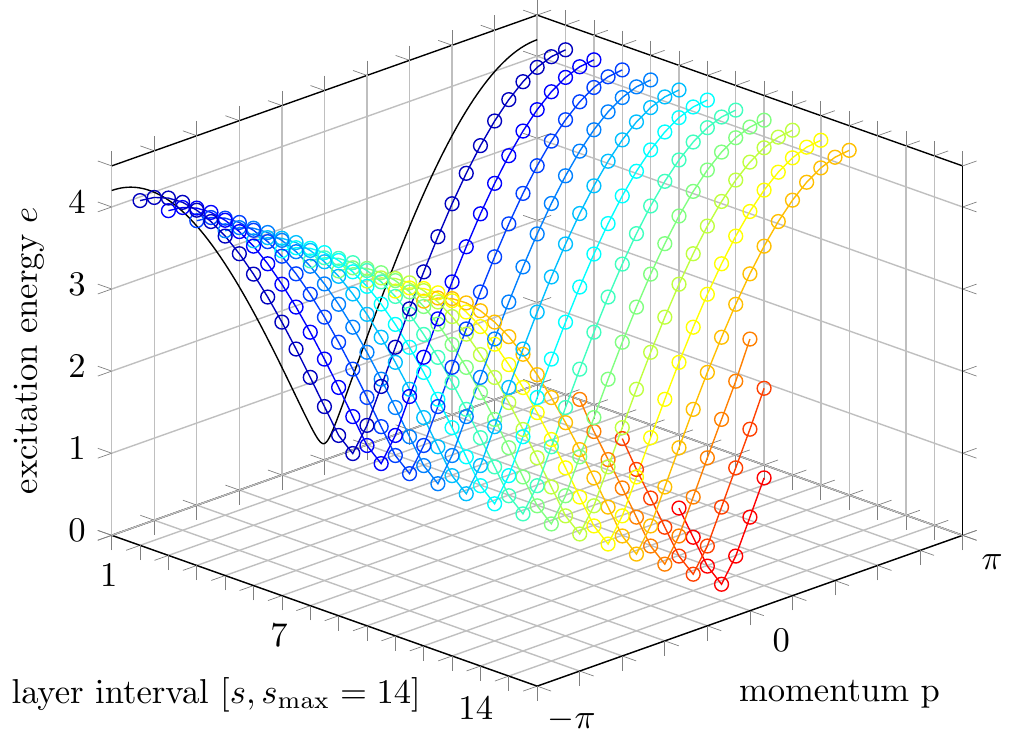}
 \caption{(Color online) Variational approximations to the quantum Ising dispersion relation $e(p)=2\sqrt{1+\lambda^2-2\lambda\cos{p}}$ (solid black line) in function of the layer index $s$ for $\lambda=1.1$, $s_{\rm max}=14$ layers and bond dimension $D=16$ using the restricted variational MPS excitation ansatz of \Fig{fig:fullexcansatz}. Momenta near the minimum $p=0$ of the dispersion relation are seen to be fully captured by restricting the variational ansatz to the top tensor(s) of the Wilson MPO, in contrast to momenta corresponding to high-energy excitations, which are observed to correspond to null states with a norm effectively approaching zero. By varying the range of the variational degrees of freedom in scale space from infrared towards the ultraviolet, we recover the high-energy excitations.} \label{fig:exc}
\end{figure}

Using the restricted variational subspace ansatz defined in \Fig{fig:fullexcansatz}, we have calculated variational approximations to the dispersion relation of the transverse Ising model \Eq{eq:ising} for all layers intervals $[s,s_{\rm max}]$, for $s=1,2,\ldots,s_{\rm max}$, of the Wilson MPO. As depicted in \Fig{fig:exc}, sweeping across layer intervals gradually allows less layers to contribute to the variational optimization. We observe that restricting the variational degrees of freedom to the top layers limits the momentum range of the ansatz to such an extent that, within the accuracy of the ground state approximation itself, momentum states corresponding to high-energy excitations can no longer be captured. Diluting exponentially on their way down the network, momentum states corresponding to high-energy excitations are seen to yield states with a norm that effectively approaches zero, leading to ill-defined generalized eigenvalue problems \Eq{eq:geig} as a function of layer index $s$ and momentum $p$. This observation of null-modes suggests an effective reduction of the Hilbert space of the momentum states corresponding to high-energy excitations at coarser renormalization group scales.

\subsection{Free fermion ansatz} \label{sec:ffresults}
We now develop a very similar approach to coarse-graining transfer matrices using free fermions. For the Ising model in \Eq{eq:ising}, we can exploit the fact that it is solvable by mapping it onto a system of free fermions to gain independent evidence in support of our ansatz \cite{fffootnote}. The additional free fermionic structure will furthermore allow us to study explicitly what happens to the transfer matrix on every coarse-graining level at the level of the ferminionic modes, which is impossible to extract from the MPO construction. To that end, we employ the modified version of the free fermionic MERA \cite{evenbly2010fermionicmera,evenbly2010bosonicmera} to effectively construct the network of coarse-graining MPOs appearing in \Eq{eq:mpoisometry}. For technical details regarding the construction, as well as further details, we refer to Appendix~\ref{app:ff}.

\begin{figure}[b]
 \centering
 \includegraphics[width=0.64\linewidth,keepaspectratio=true]{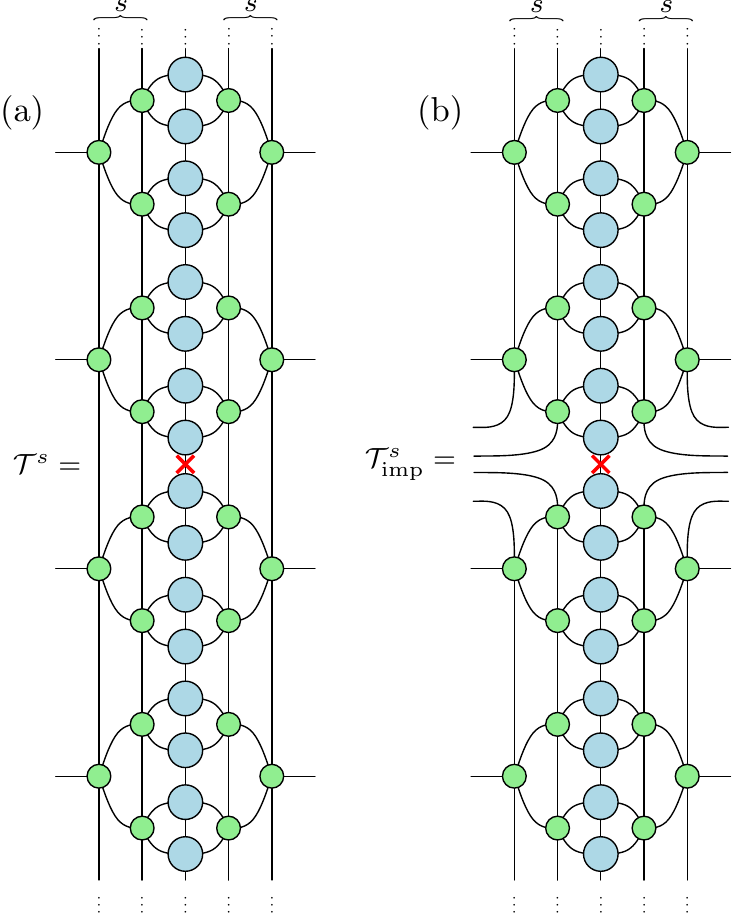}
 \caption{(Color online) (a) Coarse-grained transfer matrix and (b) partially compressed transfer matrix, obtained by application of $s$ layers of coarse-graining MPO. The same isometry is used here for the top part (i.e. ket) and the bottom part (i.e. bra) of the original MPS.}
 \label{fig:freefermiontm}
\end{figure}

\begin{figure} [t]
\begin{center}
  \includegraphics[width=\columnwidth]{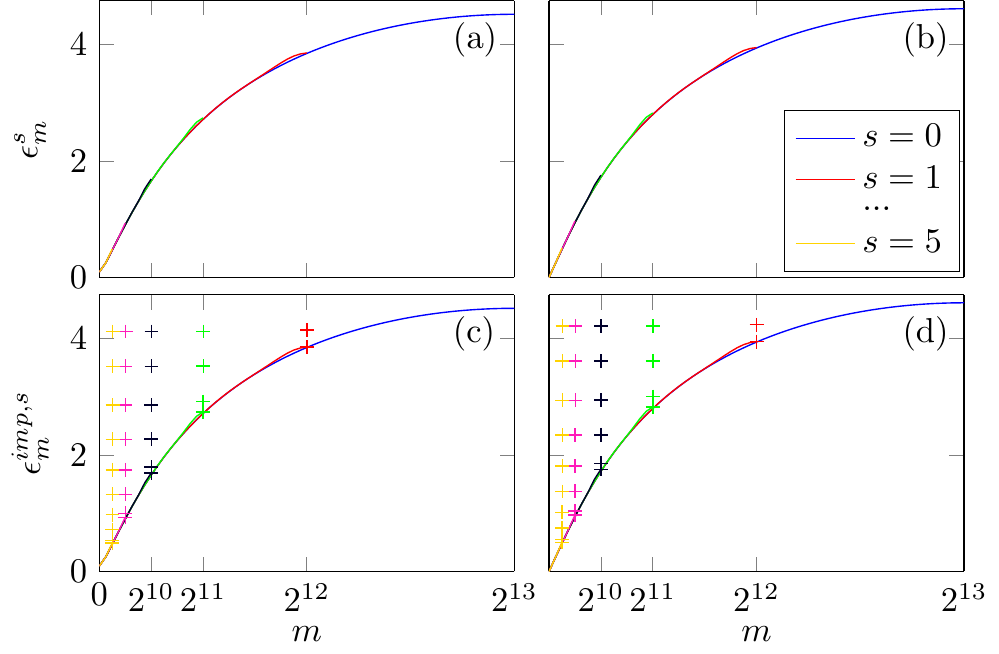} 
\end{center}
  \caption{(Color online) Single particle spectrum of (a,b) the coarse-grained transfer matrix $\mathcal{T}^s$ in \Fig{fig:freefermiontm}(a), and of (c,d) the impurity transfer matrix $\mathcal{T}^s_{\rm imp}$  in \Fig{fig:freefermiontm}(b), at different layers s. In (c,d) the approximately continuous part of the spectrum is drawn with a solid line, while pluses mark the discrete high-energy modes which are localized at, and can be associated with, the impurity degrees of freedom. Results\cite{fffootnote} for non-critical $\lambda = 1.1$ (a,c) and critical $\lambda=1$ (b,d). The initial transfer matrix was obtained by applying 4096 layers of MPOs with bond dimensions equivalent to $\chi_s = 2^1$ and $d_s = d'_s = 2^4$.} 
  \label{fig:TMSpectrum}
\end{figure}

The renormalized transfer matrix $\mathcal{T}^s$ is depicted in \Fig{fig:freefermiontm}(a). At each renormalization scale $s$ it can be diagonalized as
\begin{equation}
{ \mathcal{T}}^{s} = \exp\left[- \sum_{m} \epsilon^{s}_m c^\dagger_m c_m \right],
\label{eq:ff_spectrum}
\end{equation}
where $c_m$ are free fermionic annihilation operators, and the spectrum is fully determined by the single particle ``energies'' $\epsilon^{s}_m$.  The coarse-graining MPOs are constructed layer by layer so as to properly describe the dominant low-energy part of the spectrum of the virtual Hamiltonian $\tilde H$. As such, the free fermionic nature of the problem allows us to keep track of the spectrum of the transfer matrix at each coarse-graining step, as presented in \Fig{fig:TMSpectrum}(a,b) for non-critical and critical systems respectively.

We then use the coarse-grained MPOs obtained above to sequentially compress the transfer matrix, as depicted in \Fig{fig:freefermiontm}(b). This allows us to observe how the spectrum of the compressed transfer matrix $\mathcal{T}_{\rm imp}^s$  is gradually emerging with growing $s$, where $s=0$ marks the original transfer matrix and  $s=s_{max}$ is a fully compressed one, see \Fig{fig:ansatz}(a) for comparison. We plot the single particle energies $\epsilon_{m}^{imp,s}$, equivalent to \Eq{eq:ff_spectrum}, in \Fig{fig:TMSpectrum}(c,d) both for non-critical and critical systems. The spectra consist of a continuous part which can be attributed to still-to-be-renormalized low-energy part of the spectrum, and discrete high-energy modes corresponding to the impurity degrees of freedom. There are two such modes emerging at each new layer (for $\chi_s=2^1$, one comes from the ket and the other one from the bra). We can therefore argue that each fermionic mode, supported on the Wilson chain, represents a different, exponentially shrinking momentum-window (recall that the original problem was translationally invariant), analogous to Wilson's renormalization group picture of the impurity problem.

The spectrum of the compressed transfer matrix is then effectively given by
\begin{align}
\epsilon^{{\rm imp},s_{\rm max}}_m \approx \epsilon(k^{\rm imp}_m),
\end{align}
where $\epsilon(k)$ is the dispersion relation of the original transfer matrix $\mathcal{T}_A$ , and $k^{\rm imp}_m \sim \lambda^m$ is a logarithmic discretization of momenta resulting from the RG scheme, at least up to some infrared cutoff related to the finite length-scale inevitably appearing in the problem. 

Most importantly, exactly the same structure of the transfer matrix spectrum was observed resulting from the conventional MPS truncation procedure \cite{rams2014truncating}. Note, however, that here the parameter $\lambda$ ($\lambda=\sqrt{2}$ in \Fig{fig:TMSpectrum}) is a constant which depends on the geometry of the tensor network through the bond dimensions $d_s ,\chi_s, d'_s$, while for the standard truncation \cite{rams2014truncating} it depends non-trivially both on the bond dimension of the MPS and correlation length appearing in the problem (possibly as the result of finite entanglement scaling for the critical system). As such, we can expect that the geometry of the tensor networks limits the precision of the ansatz in a similar way to using a finite bond dimension in MPS, which we discuss in more detail in Appendix~\ref{app:ff}.

\begin{figure} [t]
\begin{center}
 \includegraphics[width=0.92\columnwidth]{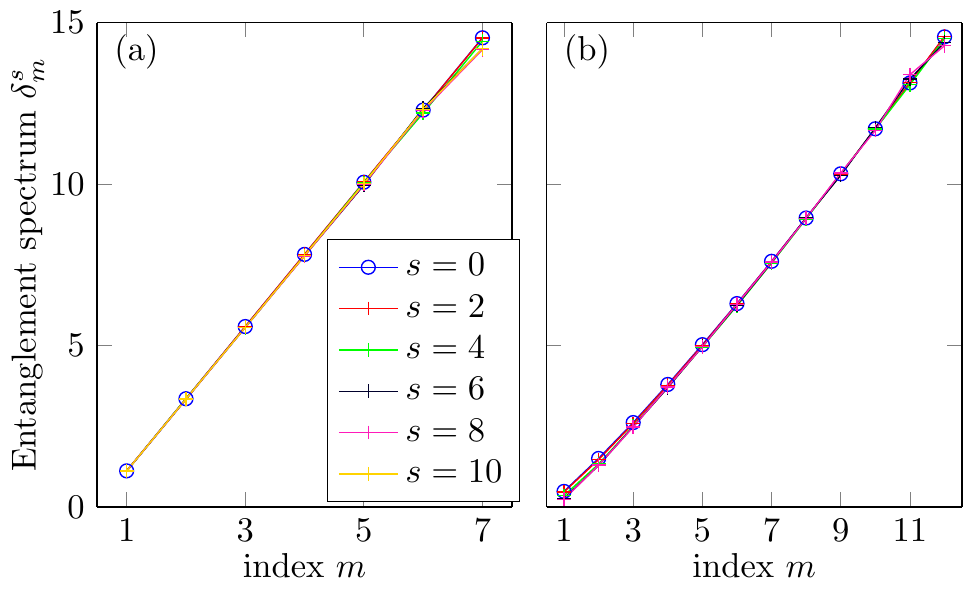} 
\end{center}
  \caption{(Color online) Dominant part of the entanglement spectrum for increasing scale $s$. The spectrum is preserved with good accuracy. Results for (a) $\lambda=1.1$ and (b) $\lambda=1$ \cite{fffootnote}. Bond dimensions equivalent to $\chi_s = 2^1$ and (a) $d_s = d'_s = 2^4$, (b) $d_s = d'_s = 2^8$. The initial transfer matrix was obtained by applying 4096 layers of MPOs. See the text for further discussion.} 
  \label{fig:EntSpectrum_ff}
\end{figure}

Finally, in order to provide direct evidence that our ansatz allows for effective compression of the transfer matrix, we show that it is reproducing the entanglement spectrum with good accuracy. To that end, the reduced density matrix of the dominant eigenvector of $\mathcal{T}_{\rm imp}^s$ can be expressed as \cite{peschel2009reduced}
\begin{equation}
\rho^s = \frac1Z \exp \left[ -2 \sum_m \delta^s_m f_m^\dagger f_m \right],
\label{eq:ent_spectrum_ff}
\end{equation}
where $f_m$ are fermionic annihilation operators, Z is the normalization factor, and the entanglement spectrum is determined by $\delta^s_m$. Consequently, the Schmidt coefficients, up to normalization, are given by $1,e^{-\delta^s_1},e^{-\delta^s_2}, e^{-(\delta^s_1+\delta^s_2)},\ldots$.
In \Fig{fig:EntSpectrum_ff} we plot the dominant part of the spectrum for growing $s$, both (a) for the non-critical system $\lambda=1.1$ and (b) the critical one $\lambda=1$.
Notice that for the non-critical case $s=s_{max}=10$ corresponds to truncating down to 10 fermionic modes, and the plot shows 7 dominant modes in the entanglement spectrum, proving that the compression is quite effective. For the critical case we stop at $s=8$, just before applying the top tensor. As encountered in \Sec{sec:mpoansatz}, simple top tensors limit the number of modes describing the state too quickly, which can significantly affect the resulting spectrum. We can further increase the precision by changing the geometry of the tensor network, either by increasing $\chi$, or equivalently, by compressing slower, which amounts to increasing $d'$ for fixed $d$, for which we refer to Appendix~\ref{app:ff}.

\section{Conclusions} \label{sec:conclusions}
In this paper, we have shown how the state compression inherent to variational MPS techniques can be interpreted as resulting from a renormalization group procedure applied to the Euclidean path integral description of a quantum system. By treating the location of the physical spin as an impurity, we were able to construct a uniform MPS representation that takes into account the degrees of freedom relevant to the impurity. The MPO structure of the coarse-graining ansatz led to a natural proposal for the structure of the MPS fixed point reduced density matrices. Furthermore, we explicitly related the different layers in the decomposition with energy scales by studying a restricted variational ansatz for excitations, which shows that perturbations at the high layers only give access to elementary excitations with a momentum near the minimum of the dispersion relation, whereas perturbations with other momenta get washed out by the subsequent layers below it. We also formulated a free fermion version of our coarse-graining ansatz that is in agreement with the results of the MPO ansatz.

Having arrived at a conceptually suggestive picture of the renormalization group structure inherent to matrix product states, we can look at the possibility of conceptual advantages in calculating correlation functions, scattering matrices, and other quantities. Further study of the behavior of excitations at this boundary between MPS and entanglement renormalization might yield insight into how to develop a proper excitation ansatz for MERA, which is even more relevant in light of its purported relation to holography. Finally, as our Wilson-based MPS furthermore suggests that the burden of entanglement can be shifted to manageable correlations between energy scales, it would be interesting to explore the possibility of continuum generalizations of our ansatz in terms of the continuous MERA \cite{haegeman2013cmera}, and to study its applicability to the numerical optimization of two-dimensional quantum lattice systems using projected entangled pair states (PEPS) \cite{verstraete2004renormalization,czarnik2015variational}.

\begin{acknowledgments}
We acknowledge inspiring discussions with Jacob Bridgeman, Glen Evenbly, Matthew Fishman, Tobias Osborne, Guifre Vidal, and especially with Micha\"el Mari\"en and Laurens Vanderstraeten. This work was supported by the Research Foundation Flanders (J.H.), the Austrian Science Fund (FWF) through grants ViCoM and FoQuS and the EC through grants QUTE and SIQS, and Narodowe Centrum Nauki (NCN, National Science Center) under Project No. 2013/09/B/ST3/01603 (M.M.R).
\end{acknowledgments}

\appendix

\section{Free fermion construction} \label{app:ff}
\subsection{Transfer matrix} 
In order to construct the tensor network representation of the ground state of the quantum Ising model in Eq. \eqref{eq:ising}, as depicted in \Fig{fig:pathintegral},  we start with the second-order Suzuki-Trotter decomposition of an operator $\ec^{-\delta H}$,
\begin{align}
 	V &= V_{1}^{1/2} V_{2} V_{1}^{1/2}, \\
      V_{1} &= \exp\left( \delta \lambda \sum_i \sigma^z_i \right) \\
      V_{2} &= \exp \left( \delta \sum_i \sigma^x_i \sigma^x_{i+1} \right),
\end{align}
for some small\cite{fffootnote} time-step $\delta$. Here, $V$ represents a single row in the two-dimensional network  in \Fig{fig:pathintegral}. It appears naturally as the transfer matrix operator in the solution of the two-dimensional classical Ising model and as such was extensively studied, see e.g. the Review \onlinecite{schultz1964ising}. Relevant for us, $V$ has a simple representation in terms of an MPO with bond-dimension $2$ (see e.g. Ref.~\onlinecite{pirvu2010matrix}), which allows us to directly obtain the full transfer matrix $\mathcal{T}_A$ \cite{rams2014truncating}(a single column in \Fig{fig:pathintegral}), of the same form as the $V$ above,
\begin{align}
	\mathcal{T}_{A} &= W_{1}^{1/2} W_{2} W_{1}^{1/2}, \\
      W_{1} &= \exp \left( K_1 \sum_{l=1}^{2L}  \tau^z_l \right) \\
      W_{2} &= \exp \left( K_2 \sum_{l=1}^{2L-1} \tau^x_l \tau^x_{l+1} \right),
\end{align}
where $\tau^{x,z}_l$ are standard Pauli matrices acting on the virtual degrees of freedom labeled with $l=1,2,\ldots 2L$, where $L$ is the number of times $V$ was applied onto the initial state (the ground state is obtained in the limit of $L \to \infty$). The physical degree of freedom is localized at the bond between sites $L$ and $L+1$. Finally, the parameters $K_{1,2}$ can be found as $K_1 = -\frac12 \log \tanh(\delta)$ and $K_2 = -\frac12 \log \tanh(\delta \lambda)$.

The transfer matrix $\mathcal{T}_A$ can be diagonalized in a standard way \cite{schultz1964ising} by mapping onto a system of free fermions with the Jordan-Wigner transformation 
$\tau_n^z = i a_{2n-1} a_{2n}$, $\tau^x_n =  a_{2n-1} \prod_{m<n} \tau^z_m$. For convenience, we introduce Majorana fermions $a_n$, $n=1,2,3, \ldots 4 L$, which are Hermitian by construction and satisfy the canonical anti-commutation relations $\left\{a_n,a_m \right\} = 2 \delta_{n,m}$. The transfer matrix can then be diagonalized as $\mathcal{T}_A = \exp \left(-\sum_{m=1}^{2L} \epsilon_m i b_{2m-1} b_{2m} \right)$, where the index $m$ can be identified with momentum and $\epsilon_m$ is the dispersion relation. The Majorana fermions $b_m = O^{\mathcal{T}_A}_{m,n} a_n$ are described by the orthogonal matrix $O^{\mathcal{T}_A}$. Since the transfer matrix has effectively open boundary conditions, some care is needed during diagonalization. To that end, we follow the procedure outlined in Ref. \onlinecite{Abraham}, obtaining $O^{\mathcal{T}_A}$ and $\epsilon_m$ numerically for some large, fixed value of $L$.

\subsection{Coarse-graining procedure} 
Our main goal is the construction of (the equivalent of) the coarse-graining MPO in Eq. \eqref{eq:mpoisometry}. To that end, the transfer matrix, which is a Hermitian, positively defined and Gaussian operator, can be described uniquely (up to a normalization) using the correlation matrix
\begin{equation}
C^{\mathcal{T}_A}_{m,n} = \frac{ \tr{ \left(a_m a_n \mathcal{T}_A\right) } }{ \tr{\mathcal{T}_A}}.
\end{equation}
We use the free fermionic MERA ansatz, adapting the construction described in \cite{evenbly2010fermionicmera} to our problem. The coarse-graining MPO is decomposed here as
\begin{align}
\vcenter{\hbox{\includegraphics[width=0.60\linewidth]{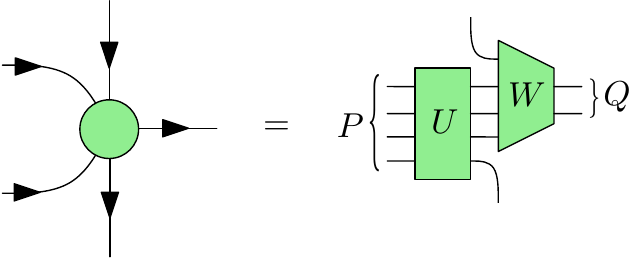}}},
\label{eq:freefermionicisometry}
\end{align}
consisting of one disentangler $U$ and one isometry $W$ in the language of MERA.  The gates $U$ and $W$, which are in principle position and layer dependent (we skip the indices to simplify notation), are parametrized in terms of SO(2P) matrices describing the local canonical transformation of Majorana fermions. They are shifted with respect to each other by one site (two Majorana modes), fixing the equivalent of the bond dimension in the coarse-graining MPO to $\chi=2^1$, with suitable boundary conditions for the tensors at the ends of the chain.
The isometries $W^\pm = W^0 Y^\pm$, where 
\begin{equation}
Y^- = \rm{diag}(\overbrace{1,1,\ldots,1,1}^{2Q},\overbrace{0,0,\ldots,0,0}^{2P-2Q}),
\end{equation}
and $Y^+ = \mathbb{I} - Y^-$ are diagonal $2P\times2P$ matrices selecting the first $2Q$ and the last $2P-2Q$ Majorana fermions, respectively. A single layer of the coarse-graining MPOs in Eq. \eqref{eq:mpoisometry} is in this picture a direct sum of disentanglers and isometries in that layer, and we will mark it as $\mathbb{W}_s^\pm$. The coarse-grained correlation matrix is obtained from the previous one as
\begin{equation}
 C^s = \mathbb{W}_s^{-\dagger} C^{s-1} \mathbb{W}_s^-,
\end{equation}
with $C^{s=0} = C^{\mathcal{T}_A}$ representing the coarse-grained transfer matrix $\mathcal{T}^s$ depicted in \Fig{fig:freefermiontm}(a).

We optimize the orthogonal matrices $U$ and $W$ in each layer, starting with the bottom layer and progressing to the top, using the standard optimization strategy for MERA \cite{evenbly2009algorithms}. The cost function is given by
\begin{equation}
 f_s(U,W) = \tr{\left( \mathbb{W}_s^{+\dagger} C^{s-1} \mathbb{W}_s^+ Y^C \right)},
\end{equation}
which allows for a local optimization. In the above, 
\begin{align}
Y^C = \bigoplus_l{ \left( \begin{matrix} 0 & -i \\ i & 0 \end{matrix} \right)}.
\end{align}
Notice that the canonical form of the correlation matrix, following a suitable canonical transformation, is
\begin{align}
C = \bigoplus_l  {\left( \begin{matrix} 1 & - i \tanh \epsilon_l/2 \\ i \tanh \epsilon_l/2 & 1 \end{matrix} \right)},
\end{align}
where $\epsilon_l$ are single particle energies of the corresponding transfer matrix. Optimizing the MERA then boils down to bringing the $\mathbb{W}_s^{+\dagger} C^{s-1} \mathbb{W}_s^+ $ correlation matrix as close as possible to the canonical form, which allows for the identification of the high-energy modes that need to be truncated during the coarse-graining. We note that the truncation procedure is equivalent to {\it tracing out} the high-energy modes in the transfer matrix, which could, in principle, introduce some bias when compared to the (more correct) {\it projection} on their ground state. We observe however that the procedure is working well, see \Fig{fig:TMSpectrum}(a,b).

\begin{figure*} [t]
\begin{center}
\includegraphics[width=0.95 \linewidth]{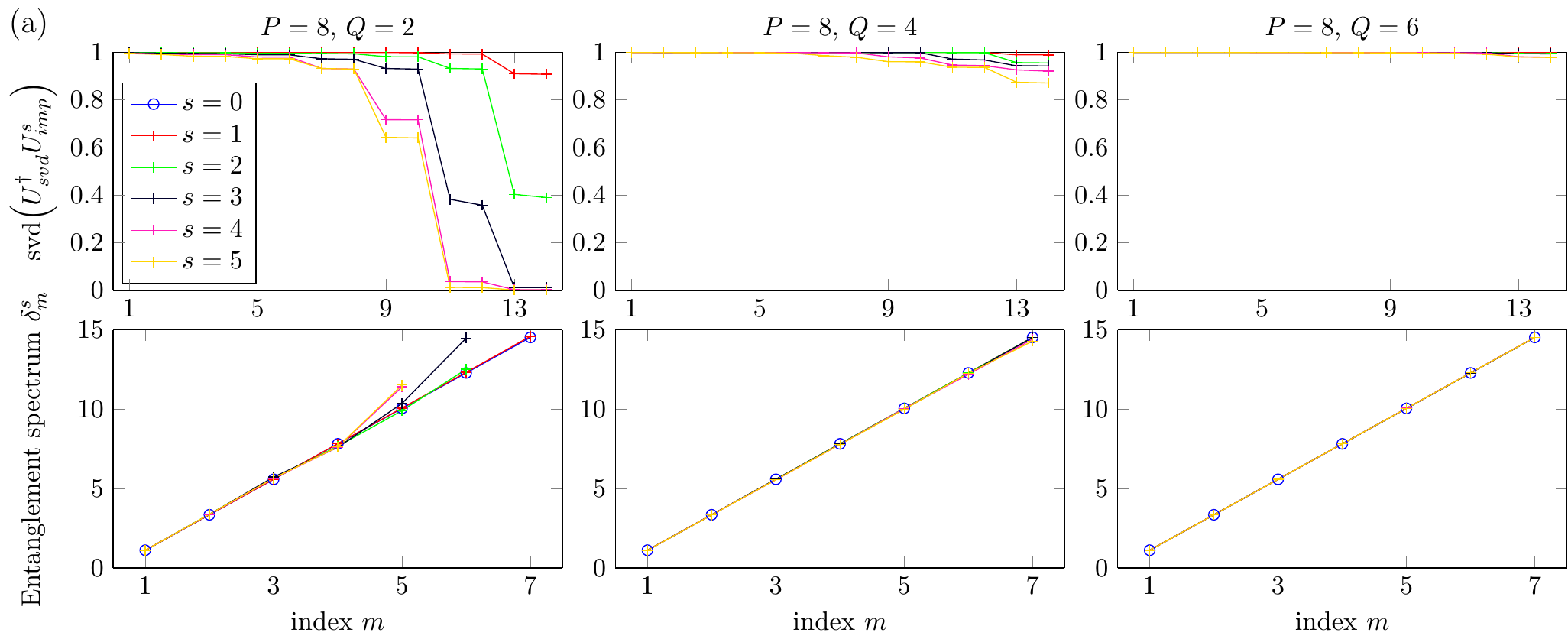} 
\includegraphics[width=0.95 \linewidth]{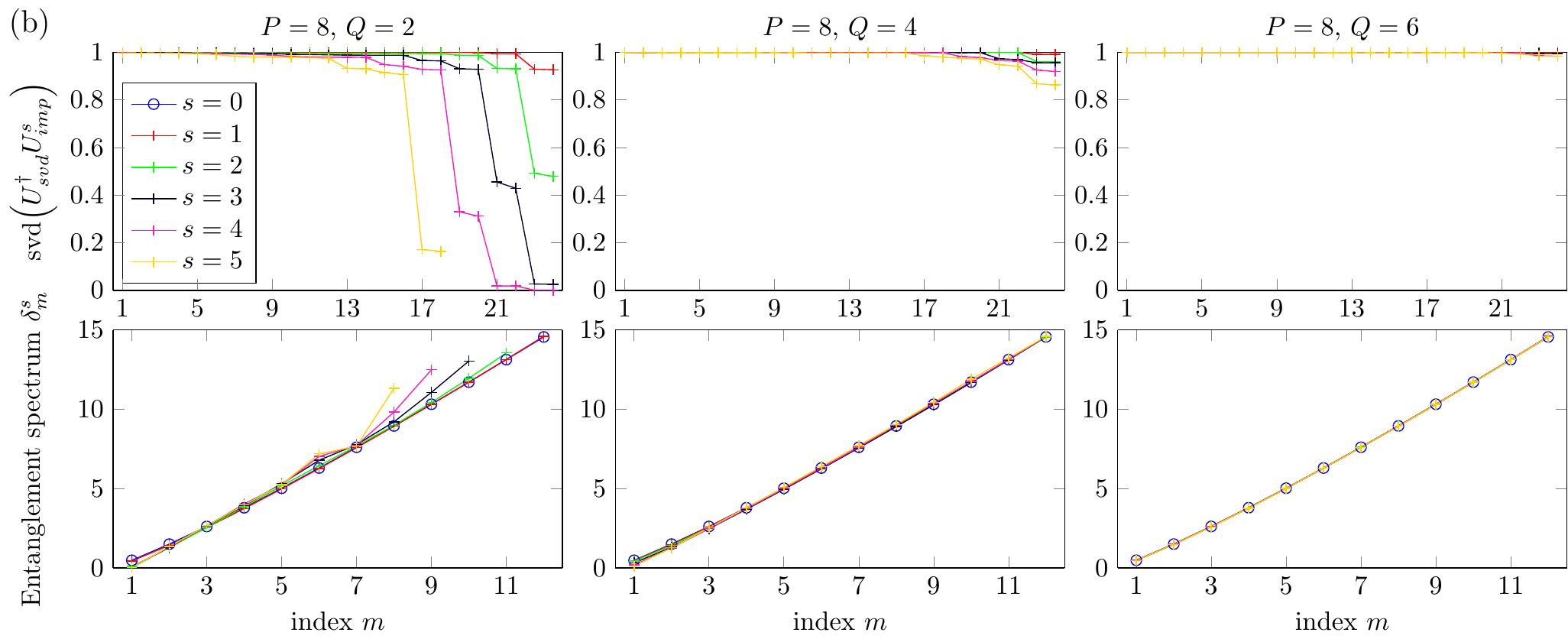} 
\end{center}
 \caption{Dominant part of the entanglement spectrum and the overlap between the isometries $U_{svd}$ and $U^s_{\rm imp}$ (see the text) for the first couple of coarse-graining layers $s$. Results for (a) $\lambda=1.1$ and (b) $\lambda=1$ \cite{fffootnote}. The initial transfer matrix was obtained by applying 4096 layers of MPOs. Comparison of different rates of compression in one layer, set by the ratio of $Q/P$, see Eq. \eqref{eq:freefermionicisometry}. This corresponds to the bond dimensions $\chi_s = 2^1$, $d_s = 2^{P/2}$ and $d'_s = 2^Q$ of the coarse-graining MPO. }
 \label{fig:impvssvd}
\end{figure*}

\subsection{Compressing the transfer matrix}
As indicated in the main text, we use the MERA generated above to construct the isometry used to compress the transfer matrix. We argue that the impurity (physical spin) can be well described using the degrees of freedom living on the boundary of its light cone, see \Fig{fig:ansatz}. To that end, we construct the isometry $U^s_{\rm imp}$ which partially compresses the transfer matrix up to the scale $s$,
\begin{align}
\vcenter{\hbox{\includegraphics[width=0.80\linewidth]{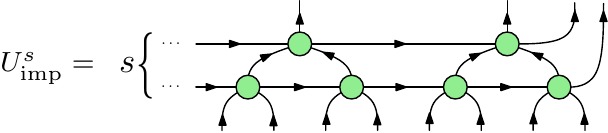}}},
\end{align}
where full compression is achieved for $s=s_{max}$. The partially compressed $\mathcal{T}^s_{\rm imp}$, described by the correlation matrix $\mathcal{C}^s_{\rm imp}$, is obtained by acting with $U^s_{\rm imp}$ on the correlation matrix $C^{\mathcal{T}_A}$ as indicated in \Fig{fig:freefermiontm}(b). The same $U^s_{\rm imp}$ is used for ket and bra of the original MPS matrix, so only about half of the tensors generated with MERA are used here in the end.

By bringing the correlation matrix $\mathcal{C}^s_{\rm imp}$ into its canonical form, we can find the spectrum of the corresponding transfer matrix as
\begin{equation}
{ \mathcal{T}}^{s}_{\rm imp} = \exp\left[- \sum_{m} \epsilon^{imp,s}_m c^\dagger_m c_m \right],
\end{equation}
in its diagonal base of Dirac fermions $c_m$. The spectrum is plotted in \Fig{fig:TMSpectrum}(c,d) and discussed further in the main text. The entanglement spectrum, Eq. \eqref{eq:ent_spectrum_ff}, which is given by the reduced density matrix  of the dominant eigenvector of $\mathcal{T}_{\rm imp}^s$, can be calculated from $\mathcal{C}^s_{\rm imp}$ as well, see e.g. Ref. \onlinecite{peschel2009reduced}.

\subsection{Comparison of impurity picture with standard truncation}
Finally, we can now directly compare the isometry $U_{\rm imp}^s$ obtained with MERA, with the standard truncation procedure described by the isometry $U_{svd}$, a $2L\times 2\chi$ matrix where $2^\chi$ is the bond dimension of the truncated MPS. The matrix $U_{svd}$ is obtained by calculating the reduced density operator of the MPS on a half-infinite chain (our initial $A^i$ in \Fig{fig:pathintegral}) and finding dominant modes in its diagonal basis \cite{rams2014truncating}. To that end, we look at the singular values of $U_{svd}^\dagger U^s_{\rm imp}$, which directly show how well the dominant modes in the entanglement spectrum are preserved during the compression.

The results for the gapped and critical cases are plotted in Figs.~\ref{fig:impvssvd} (a) and (b) respectively, together with the resulting entanglement spectra. Notice that each mode in the entanglement spectrum corresponds to two Majorana modes in term of the isometries $U$. We show the results for the first couple of layers, up to $s=5$, and various ratios of $Q/P$, see Eq. \eqref{eq:freefermionicisometry}. This ratio sets an effective light cone, and determines what fraction of the large energy modes of the original transfer matrix is renormalized during each coarse-graining step. Notice that from that perspective, one step for $Q/P = 2/8$ is equivalent to two steps for $Q/P=4/8$ and almost five steps for $Q/P=6/8$. At the same time, the  number of modes describing the impurity resulting from each step is the same. The trade-off between the rate of compression and the accuracy can be readily seen. As already described in the main text,  we expect that this is directly equivalent to obtaining MPS with given bond dimension $D$ using the standard procedure. For instance, for $Q/P=2/8$ only the first few modes are described accurately corresponding to smaller $D$, while for $Q/P=6/8$ all the dominant (as plotted) modes are reproduced corresponding to large $D$ (we should be able to obtain the same result by increasing $\chi$). The choice of $Q/P=4/8$ (i.e., $d=d'$) seems to be a good compromise between the two.

\section{Isometric MPO versus MERA} \label{app:mera}
In the main text, we have proposed to characterize an isometric matrix product operator (MPO) locally by introducing a five-legged tensor $g$, which can be interpreted as a map $g : \mathbb{I} \otimes \mathbb{I} \otimes \mathbb{V} \to \mathbb{O} \otimes \mathbb{V}$, where $\mathbb{I}$, $\mathbb{V}$, and $\mathbb{O}$, respectively, refer to the vector spaces of the incoming indices, the virtual indices and the outgoing indices. As a variational set of states, the ansatz \Eq{eq:tmansatz}, constructed of layers of isometric MPOs, includes the MERA. Explicitly, the ansatz \Eq{eq:tmansatz} encompasses MERA in the sense that when the local isometry $g$ is, for example, interpreted as a disentangler $u$ and an isometry $w$ of a binary MERA,
\begin{align}
\vcenter{\hbox{
\includegraphics[width=0.5\linewidth]{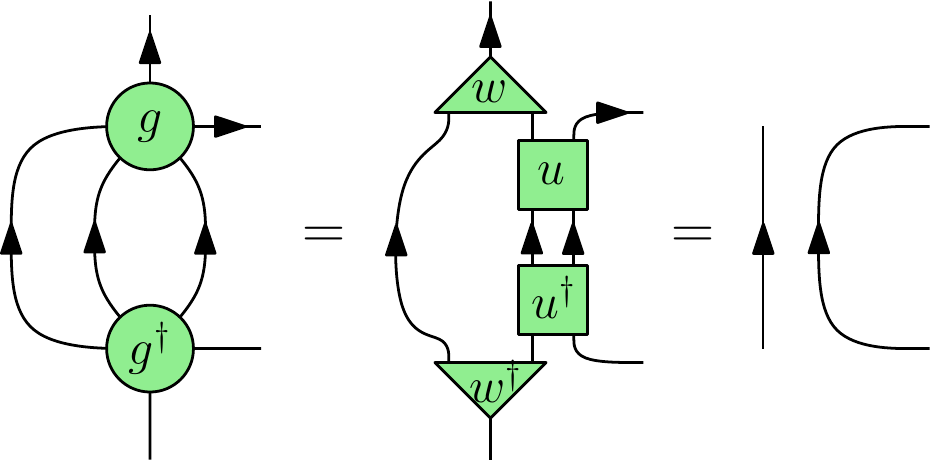}}}, \label{eq:appisoleft}
\end{align}
\begin{align}
\vcenter{\hbox{
\includegraphics[width=0.5\linewidth]{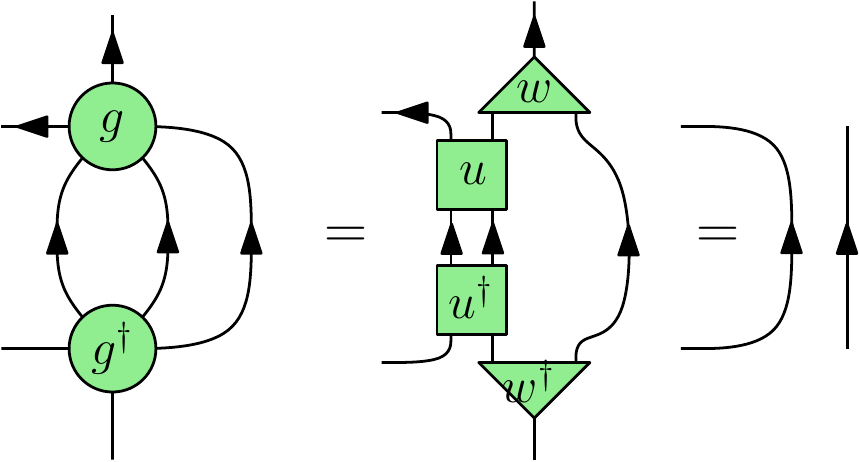}}}, \label{eq:appisoright}
\end{align}
the internal bond connecting the disentangler and the isometry is in no way restricted by $d=\dim (\mathbb{I})$. Note, however, that there is no local gauge transformation acting purely on the virtual level of the coarse-graining MPO, which can transform between Eqs.~\eqref{eq:appisoleft} and \eqref{eq:appisoright}. Intuitively, we expect that the virtual dimensions of the MPO structure within every layer result in a coarse graining scheme that is quasi-local rather than strictly local as in the MERA). This difference is due to the different causal cone structures. The causal region of the MPO ansatz is potentially larger than that of MERA as it, at most, extends towards infinity in one of both directions. The MERA causal cone, which is strictly local, is included in this extended causal cone. The more general MPO ansatz still enables us to renormalize translation invariant MPOs into translation invariant MPOs and renders the identification of the Wilson chain very straightforward.

\bibliography{mpsrg}

\end{document}